\documentclass[a4paper, reprint, twoside,
  nofootinbib,longbibliography]{revtex4-1}
\makeatletter
\def\input@path{{fig/}} \def\p@figure{Fig.~}
\makeatother
    \usepackage[utf8]{inputenc}
    \usepackage[english]{babel}
    \usepackage[x11names]{xcolor}
    \usepackage{geometry}
    \usepackage{graphicx}
    \usepackage{float}
    \usepackage{subcaption}
    \usepackage[inline]{enumitem}
        \setlist[enumerate,1]{%
            label=\textsuperscript{\arabic*},
        }
        \newlist{inlinelist}{enumerate*}{1}
        \setlist*[inlinelist,1]{%
            label=set\textsuperscript{\arabic*}\!\!,
        }
    \usepackage{appendix}
    \usepackage{mathtools}
        \allowdisplaybreaks
    \usepackage{bm}
    \usepackage{amssymb}
    \usepackage{amsthm}
        \let\vec\bm

        \DeclareMathOperator{\EX}{\mathbb{E}}
        \DeclarePairedDelimiter\ton{(}{)}
        \DeclarePairedDelimiter\qua{[}{]}

        \DeclarePairedDelimiter\norm{\lVert}{\rVert}
        \DeclarePairedDelimiter\mean{\langle}{\rangle}
    \usepackage{hyperref}
        \hypersetup{
            colorlinks=true,
            citecolor=black,
            linkcolor=black,
            urlcolor=blue}
        \let\originaleqref\eqref
        \renewcommand*{\eqref}[1]{Eq.~\originaleqref{#1}}
        \makeatletter
            \def\p@subfigure{Fig.~\thefigure}
        \makeatother

\begin{document}
\title{The effect of Collaborative--Filtering based Recommendation Algorithms\\ on Opinion Polarization}
\author{Alessandro Bellina$^{1, 2}$}
\author{Claudio Castellano$^{3, 2}$}
\author{Paul Pineau$^{4}$}
\author{Giulio Iannelli$^{2, 5}$}
\author{Giordano De Marzo$^{1, 2, 6, 7}$}
\affiliation{$^1$Dipartimento di Fisica Universit\`a ``Sapienza”, P.le A. Moro, 2, I-00185 Rome, Italy.}
\affiliation{$^2$Centro Ricerche Enrico Fermi, Piazza del Viminale, 1, I-00184 Rome, Italy.}
\affiliation{$^3$Istituto dei Sistemi Complessi (ISC-CNR), Via dei Taurini 19, I-00185 Roma, Italy}
\affiliation{$^4$\'Ecole Normale Sup\'erieure Paris-Saclay, 4 Avenue des Sciences, 91190 Gif-sur-Yvette, France.}
\affiliation{$^5$Dipartimento di Fisica, Università di Roma “Tor Vergata”, 00133 Roma, Italy.}
\affiliation{$^6$Complexity Science Hub Vienna, Josefstaedter Strasse 39, 1080, Vienna, Austria.}
\affiliation{$^7$Sapienza School for Advanced Studies, ``Sapienza'', P.le A. Moro, 2, I-00185 Rome, Italy.}
\date{\today}
\begin{abstract}
    A central role in shaping the experience of users online is played
    by recommendation algorithms. On the one hand they help retrieving
    content that best suits users taste, but on the other hand they
    may give rise to the so called ``filter bubble'' effect, favoring
    the rise of polarization. In the present paper we study how a
    user--user collaborative--filtering algorithm affects the behavior
    of a group of agents repeatedly exposed to it. By means of
    analytical and numerical techniques we show how the system
    stationary state depends on the strength of the similarity and
    popularity biases, quantifying respectively the weight given to
    the most similar users and to the best rated items. In particular,
    we derive a phase diagram of the model, where we observe three
    distinct phases: disorder, consensus and polarization. In the
    latter users spontaneously split into different groups, each
    focused on a single item. We identify, at the boundary between the
    disorder and polarization phases, a region where recommendations
    are nontrivially personalized without leading to filter
    bubbles. Finally, we show that our model well reproduces the
    behavior of users in the online music platform last.fm. This
    analysis paves the way to a systematic analysis of recommendation
    algorithms by means of statistical physics methods and opens to
    the possibility of devising less polarizing recommendation
    algorithms.
\end{abstract}
\maketitle
\section{Introduction}
The growth of polarization and
radicalization observed in recent years \cite{Pew2017,
  chitra2020analyzing, maes2015will} is a phenomenon that can
potentially undermine the functioning and stability of democratic
societies. In this context, the critical role played by online
platforms has been widely recognized \cite{nguyen2014exploring,
  bryant2020youtube, o2013extreme}, but the detailed mechanisms by
which exposure to online content drives polarization at the population
level are still to be fully clarified. Recommendation algorithms, together with more traditional media such as television \cite{muise2022quantifying}, are
believed to be among the key factors, since they strongly influence
users online experience by selecting, based on past behavior,
the new information users are exposed to
\cite{linden2003amazon,lu2012recommender,smith2017two}. Such
algorithms are fundamental for filtering and selecting the content we
are interested to, a sorely needed task, given the overwhelming amount
of information available online. On the other side, however,
recommendation algorithms produce a feedback loop that naturally tends
to bias future choices, reducing the diversity of available content
and thus favoring the so called ``filter-bubble'' effect and the
consequent polarization of opinions \cite{pariser2011filter,
  dillahunt2015detecting, nagulendra2014understanding,
  gravino2019towards, kirdemir2021exploring}. Filter bubbles,
occurring when users are mainly exposed to news and content aligned to
their beliefs, are similar to the much investigated ``echo
chambers'' \cite{Cinelli2021, Cota2019, barbera2015tweeting}. While
the latter result from homophylic interactions among users, which tend
to interact with people sharing their same opinions, the former are
produced by algorithmically biased recommendations in online
platforms.\par

Recommendation algorithms are widely used by most of the websites we
visit everyday, examples being ``suggested for you'' posts on
Facebook, recommended items on the Amazon online shop or Google
personalized PageRank. Such algorithms are designed to allow easy
access to content we are expected to be interested in so as to
maximize our engagement with the platform. Collaborative--filtering
\cite{herlocker2000explaining,su2009survey} is a paradigmatic approach
to algorithmic recommendation which, despite its simplicity, is
employed by online giants such as Amazon \cite{lu2012recommender,
  smith2017two}. The underlying principle is that past behavior of
users can be exploited to determine the similarity between them or
between items, which can then be used to identify new content users
will most likely appreciate. The first case corresponds to
``user--user'' collaborative--filtering, while the latter to the
``item--item'' one. In the following we will use interchangeably the
terms ``item'' or ``opinion'' as equivalent ways to refer to a generic
piece of content available on the platform.\par

In the past years much attention has been devoted to the study of how
microscopic interactions among users shape collective phenomena at the
population level \cite{Castellano2009,Sen2014} and in particular to
the investigation of how the polarization of opinions emerges from
such interactions \cite{Ciampaglia2018,Sirbu2019,
  freitas2020imperfect,Baumann2020,Baron2021}. The effect of
recommendation algorithms has received less attention and only
recently scholars started to model their interplay with the dynamics
of opinions. For instance Refs. \cite{DeMarzo2020, iannelli2022filter}
approached the problem by endowing a voter model with an external
field which represents users interaction with their past history, thus
mimicking content recommendation. A similar methodology has been
proposed in Ref. \cite{Perra2019}, where the effect of recommendations
based on agents' present state is considered, while in other works
\cite{Peralta2021,peralta2021opinion} the effect of the recommendation
algorithm is modeled by filtering the interactions between an
individual and its neighbors, depending on their state. In all cases
the conclusion is that recommendation algorithms may play a crucial
role in enhancing opinion polarization and fragmentation.
Also the effect of link recommendations (algorithms suggesting new
social connections) has been analyzed, revealing that personalized
suggestions of new friends can increase
polarization~\cite{santos2021link,cinus2022effect, valensise2022dynamics}
and favor inequality and biases~\cite{ferrara2022link, espin2022inequality}.\par

All the studies concerning content recommendations, despite providing
useful insight on the possible effects of their implementation, consider
exceedingly simple algorithms, coupled to highly stylized opinion
dynamics models. As a consequence they do not shed much light on the
effect of realistic recommendation algorithms adopted by online
platforms. In order to fill the gap between theoretical modeling and
real implementations, in this paper we present a systematic
study of a model for user--user collaborative--filtering.
We find that, depending on two
parameters (the strength of the similarity bias $\alpha$ and of the
popularity bias $\beta$, see below for definitions), the system can be
in three different phases: disorder, consensus and polarization. In
particular, when the two biases are sufficiently large, the system
undergoes a spontaneous breaking of users and items symmetry, leading
to the formation of polarized groups and giving rise to the filter
bubble effect. Such a drawback can be avoided at the boundary between
disorder and polarization, where the algorithm provides meaningful
recommendations without inducing opinion polarization.
Finally, we use our model to reproduce the behavior of users in
the online music platform last.fm, determining, within our
modeling framework, the strength of
the similarity and rating biases they are subject to.\par

These results show that a statistical physics approach to
recommendation algorithms is crucial in understanding their effect on
opinion polarization, while also being a powerful tool for determining
the best parameters to be used in their implementation.
\section{Definition of the model}
Let us consider a system composed of $N$ users which iteratively
choose (click) among $M$ items or opinions. We denote by $U$ (with
$|U|=N$) the set of all users, while $I$ (with $|I|=M$) is the set of
items. At time $t$, each user $u$ is described by a $M$-dimensional
vector $\vec{r}_u(t)=\{r_{u1}(t), \dots, r_{uM}(t)\}$ whose components
$r_{ui}(t)$ are given by the number of times user $u$ has clicked on item
$i$ so far.
In the following we refer to the $r_{ui}$ as ratings and we
assume that clicking on an item expresses (positive) interest in it.
Initially all ratings are set equal to $r_0$, i.e. $r_{ui}(0)=r_0$ for
all users $u$ and items $i$.  These initial conditions reflect the
absence of any a priori knowledge about users' taste and mimic the so
called ``cold start'' of recommendation algorithms
\cite{schein2002methods}. Note that in real systems the number of
items available to users is typically enormous. For instance there are
almost 100 million tracks on Spotify and around 350 million products
are available on Amazon Marketplace. As a consequence in the following
we will be interested in taking the large $M$ limit. In order to do
so, as explained in Appendix \ref{app:scaling_initial_conditions}, we
have to set $r_0\sim M^{-1}$; in the following we take
$r_0=1/(M-1)$ unless specified otherwise.\par

At each time step, of duration $\delta t=1/N$, a user $u$ is selected
at random and he/she clicks on item $i$ with probability $R_{ui}(t)$
\begin{align*}
    R_{ui}(t)&=P(r_{ui}(t+\delta t)=r_{ui}(t)+1)=\\
        &=\text{Prob($u$ is selected at time $t$ and clicks on $i$)}.
\end{align*}
The specific form of this normalized probability
($\sum_{u,i} R_{ui}(t)=1$) encodes how the recommendation algorithm affects the
user behavior. The idea behind the collaborative--filtering mechanism
is that such a probability should be the larger the more item $i$ is
positively rated by users similar to $u$. More precisely, we quantify
the similarity $s_{uv}$ between users $u$ and $v$ as the cosine
similarity of their rating vectors, that is
\[
    s_{uv}=\frac{\vec{r}_u \cdot
      \vec{r}_v}{\norm{\vec{r}_u}\norm{\vec{r}_v}}=\dfrac{\sum_i
      r_{ui} r_{vi}}{\sqrt{\sum_i r_{ui}^2} \sqrt{\sum_i r_{vi}^2}}.
\]
In these terms we define the transition probability $R_{ui}(t)$ as 
\begin{equation}
 R_{ui}(t)=\frac{1}{N}\sum_{v=1}^N \dfrac{s_{uv}^{\alpha}(t)}{\sum_w s_{uw}^{\alpha}(t)} \dfrac{r_{vi}^{\beta}(t)}{\sum_j r_{vj}^{\beta}(t)}.
 \label{eq:transition_prob}
\end{equation}
The two parameters $\alpha$
and $\beta$ quantify the strength, respectively, of the similarity
bias and of the popularity bias. Indeed, the larger $\alpha$, the more
users are biased toward items liked by the agents they are more
similar to, while the larger $\beta$ the more users are biased toward
items already selected in the past. \eqref{eq:transition_prob} is a direct generalization of the standard user-user collaborative filtering expression \cite{resnick1994grouplens, lu2012recommender}. The latter corresponds to $\alpha=\beta=1$ and reads
\[
	\text{Score}_{ui}=\sum_v^Ns_{uv}\frac{r_{vi}}{\sum_jr_{vj}}.
\]
The clicking probability of a user is then obtained normalizing the scores by their sum.

Note that with the present definition the probability for user $u$ is
affected by the ``self--interaction'' with his/her past, as the sum
includes the term weighted by $s_{uu}=1$.
Because of self--interaction,
when $\alpha \to \infty$ each agent interacts only with him/herself
and users are completely independent. The framework we are considering
corresponds to a collaborative--filtering with implicit feedback,
meaning that the appreciation users give to items is not directly
available, but rather it is derived by the number of times users click
on items. This situation occurs, for instance, in music streaming
platforms, where $r_{ui}$ corresponds to the number of times user $u$
listened to song (or artist) $i$.
\section{Behavior for limit values of the parameters}
By inspecting \eqref{eq:transition_prob} it is easy to anticipate
that, depending on the strength of the similarity and popularity
biases, the system can show very different behaviors. In particular,
three distinct phases can be identified by considering simple limits.
\begin{itemize}
    \item $\beta=0$, $\forall \alpha$: \textbf{Disorder}\\
        Without a popularity bias, \eqref{eq:transition_prob} reduces to
        \[
            R_{ui}=\frac{1}{N}\sum_v \dfrac{s_{uv}^{\alpha}}{\sum_w s_{uw}^{\alpha}} \dfrac{1}{M}=\dfrac{1}{NM}
        \]
        and thus all users behave as random clickers. In this case all
        the items share the same probability of being clicked and the
        system is \textit{disordered}, meaning that any user rates
        equally (on average) any item.

      \item $\alpha=0$ and $\beta=\infty$: \textbf{Consensus} \\
        When the similarity bias is set to $\alpha=0$ the
        transition probability simply is
        \[
            R_{ui}=\dfrac{1}{N^2} \sum_v \dfrac{r_{vi}^{\beta}}{\sum_j r_{vj}^{\beta}}
        \]
        and is independent of the user $u$. Moreover, since the
        popularity bias is maximal, it holds ${r_{vi}^{\beta}}/{\sum_j
          r_{vj}^{\beta}}= \delta_{i, i_v}$, where $i_v$ is the most
        rated item by user $v$. As a consequence, denoting as
        $N_i=\sum_v\delta_{i, i_v}$ the number of users having $i$ as
        the most rated item, we get
        \[
            R_{ui}=\frac{1}{N^2}\sum_v\delta_{i, i_v}=\frac{N_i}{N^2}.
        \]
        This expression implies that users are more likely to click on
        the globally most popular item, thus originating a feedback
        loop which for large time is expected to make such an item
        become the most rated for each user. This means that the
        system evolves toward a \textit{consensus} phase, where all
        users agree on the same opinion. In this consensus phase the
        recommendation algorithm always suggests the same item to all
        users.

      \item $\alpha=\infty$ and $\beta=\infty$: \textbf{Polarization}
        \\ In this case both the rating and the similarity biases are
        maximal and \eqref{eq:transition_prob} becomes
        \[
            R_{ui}=\frac{1}{N} \delta_{i,i_u},
        \]
        where $i_u$ is the opinion more frequently clicked on by user
        $u$. As a consequence users persistently stick to their first
        random choice (note that at $t=0$ all ratings are the same,
        and so all items have the same probability to be chosen),
        giving rise to \textit{polarization} and to the filter bubble
        effect. Indeed in the polarized phase the recommendation
        algorithm suggests to each user just one specific item, but
        such an item varies from user to user.
\end{itemize}

In order to investigate the model, it is useful to introduce the normalized ratings $\hat{r}_{ui}$, defined as
\begin{equation}
    \hat{r}_{ui}(t)=\frac{r_{ui}(t)}{\sum_i r_{ui}(t)}\approx \frac{r_{ui}(t)}{t+Mr_0},
        \label{eq:normrat}
\end{equation}
where the approximation comes from the assumption that in a time
interval $\Delta t=N\cdot\delta t=1$ each user is updated once on
average. The normalized ratings satisfy, for asymptotically large
times, the Martingale Property, i.e. $\EX[\hat{r}_{ui}(t+\delta
  t)]=\EX[\hat{r}_{ui}(t)]$. Since they are also limited in $(0,1)$,
this ensures that these random variables converge to an asymptotic
limit for large times. Thus they are the right variables to look at in
order to find asymptotic stationary solutions of the system. By looking at the evolution of the normalized ratings $\hat{r}_{ui}$ for various values
of the biases we can observe the different phases identified by inspecting the behavior of the model in the limit cases discussed above.
\begin{figure*}
  \centering
  \begin{subfigure}[b]{0.475\textwidth}
    \centering
    \includegraphics[width=\textwidth]{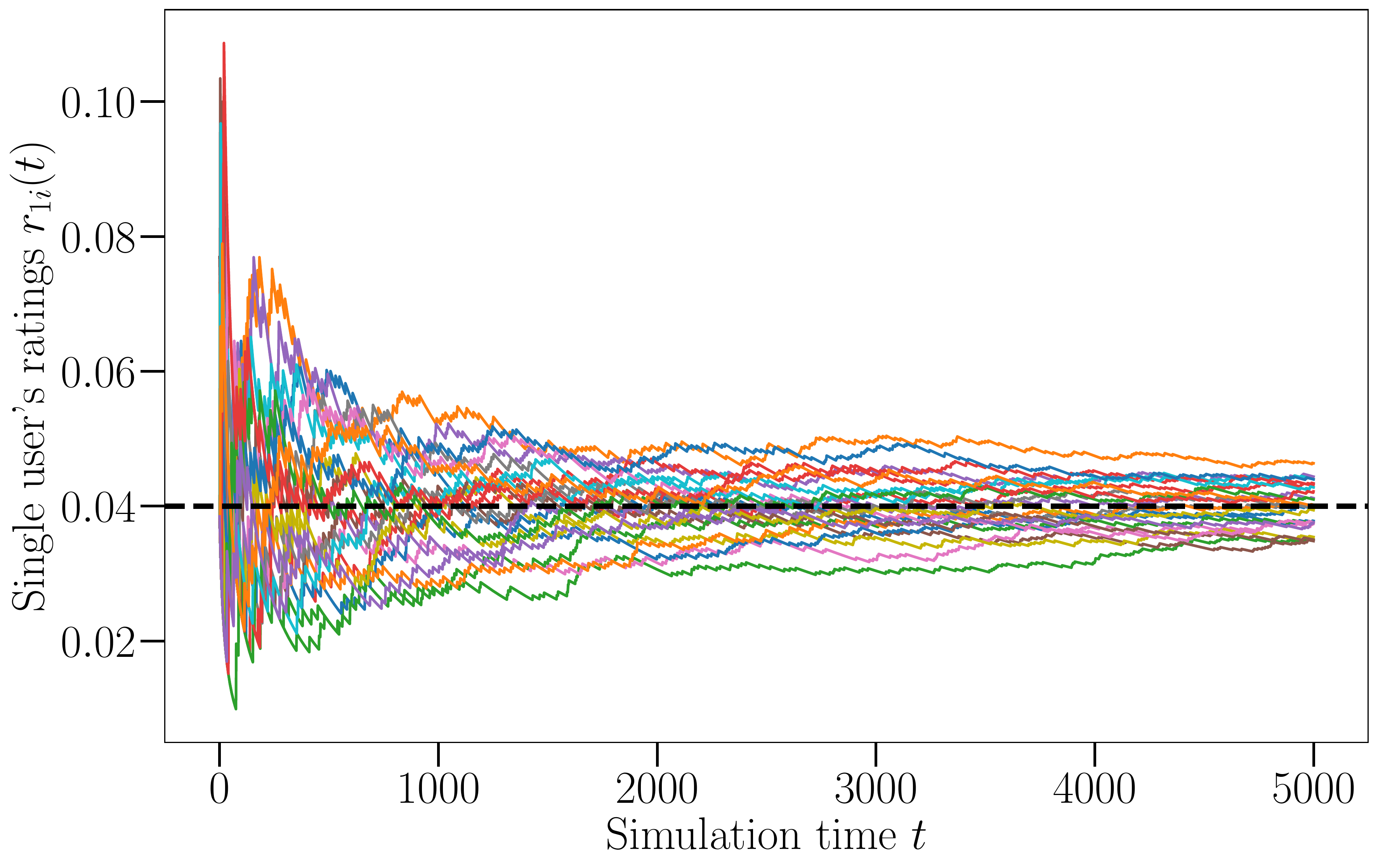}
    \caption{Ratings of user 1 on different items. $N=M=25$, $\alpha=0.5$ and $\beta=0.5$.}     
    \label{fig:phen_disorder}
  \end{subfigure}
  \hfill
  \begin{subfigure}[b]{0.475\textwidth}  
    \centering 
    \includegraphics[width=\textwidth]{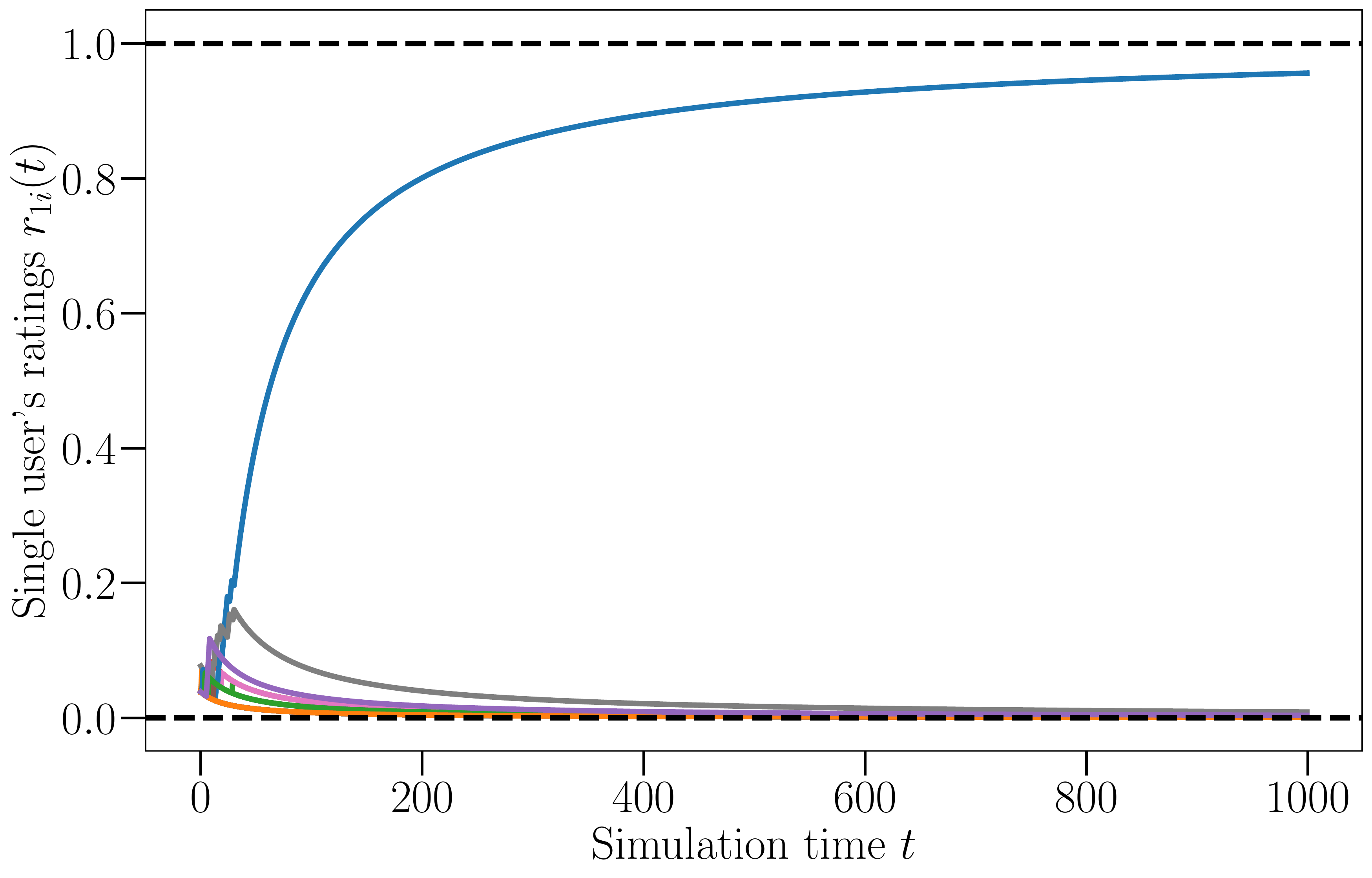}\\
    \caption{Ratings of user 1 on different items. $N=M=25$, $\alpha=0.5$ and $\beta=5$.}    
    \label{fig:phen_cons_item}
  \end{subfigure}
  \begin{subfigure}[b]{0.475\textwidth}   
    \centering 
    \includegraphics[width=\textwidth]{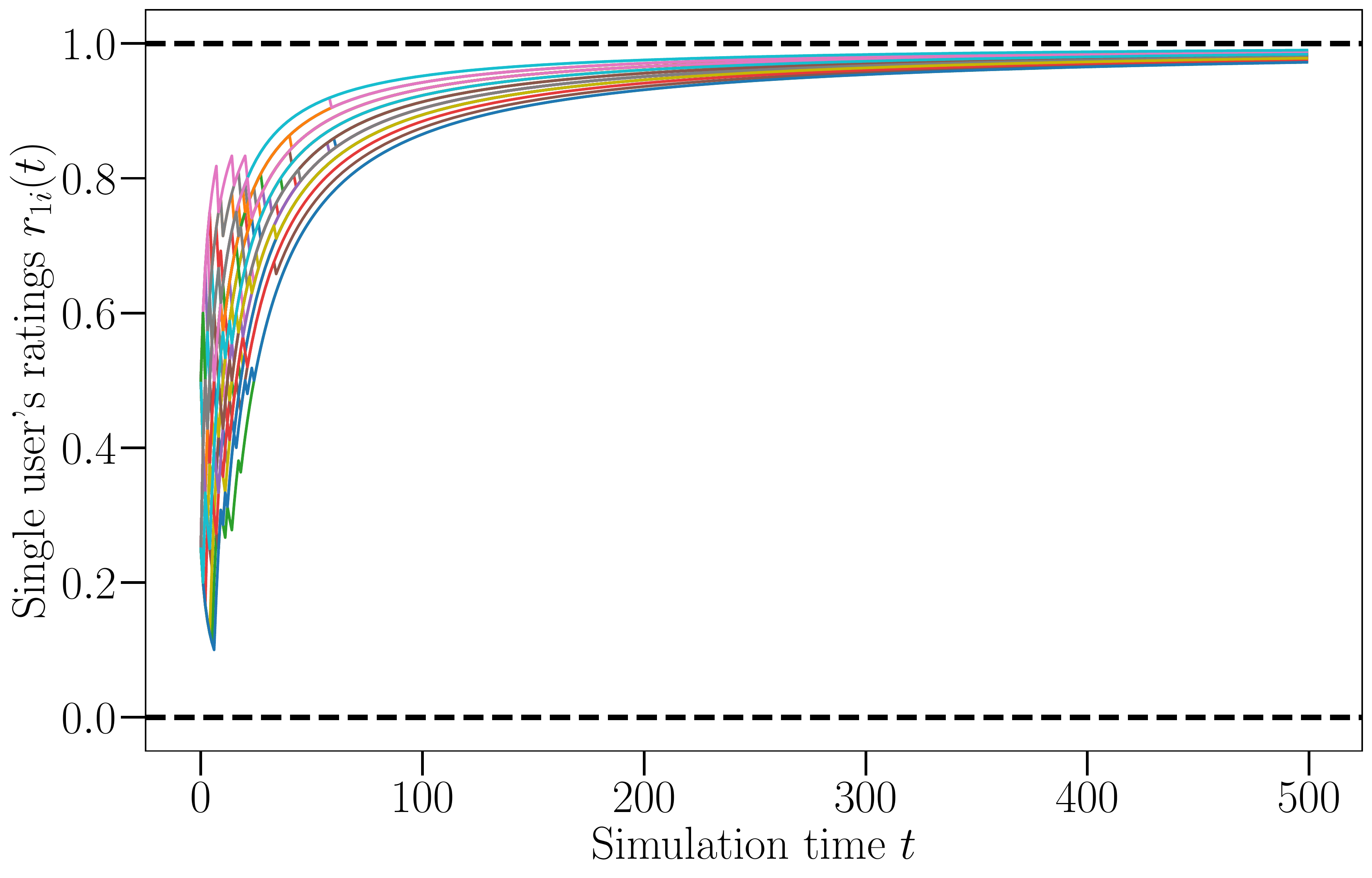}
    \caption{Ratings for item 1 of different users. $N=M=40$, $\alpha=0.5$ and $\beta=5$.}      
    \label{fig:phen_cons_user}
  \end{subfigure}
  \hfill
  \begin{subfigure}[b]{0.475\textwidth}   
    \centering 
    \includegraphics[width=\textwidth]{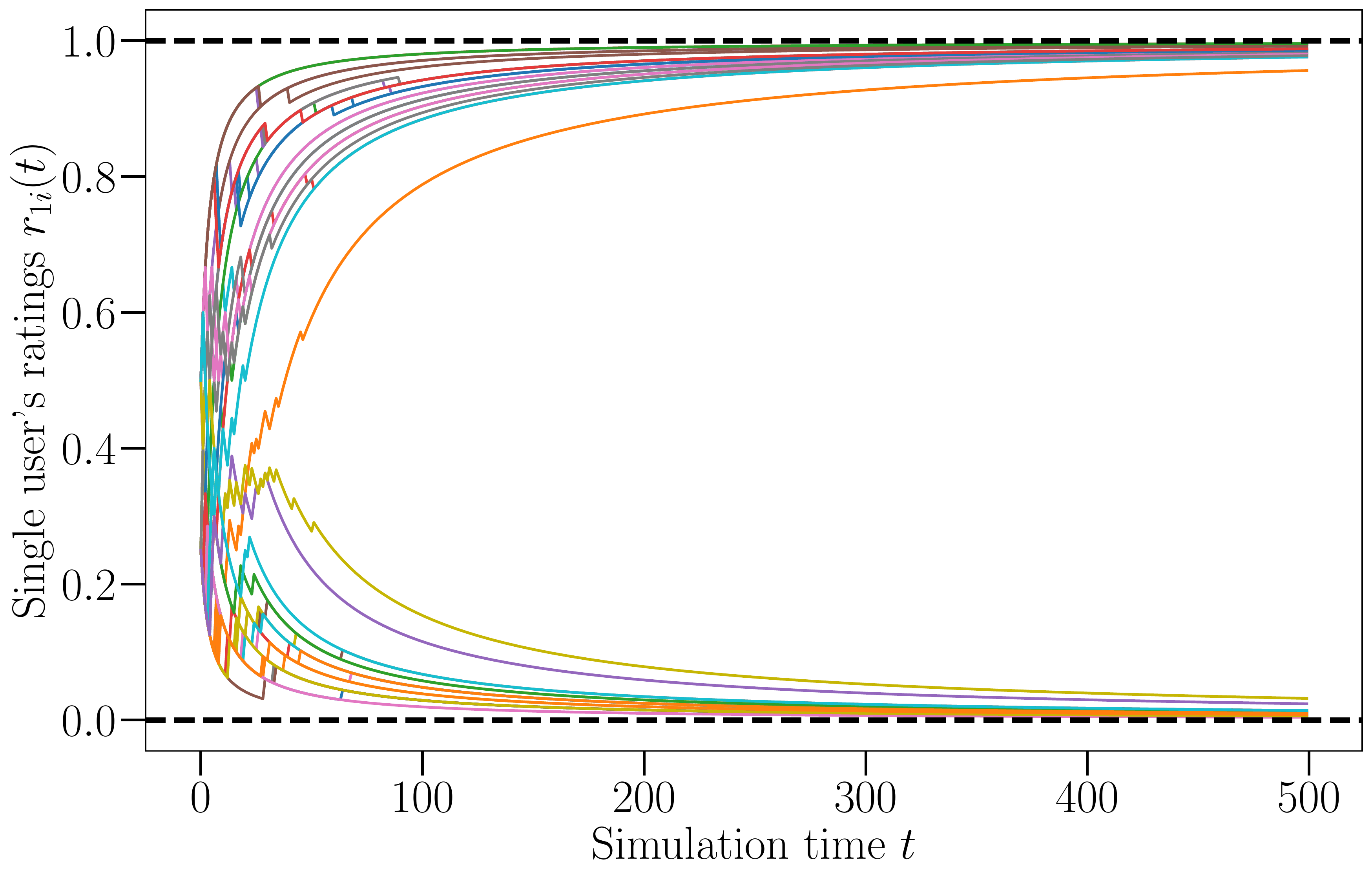}
    \caption{Ratings for item 1 of different users. $N=M=40$, $\alpha=5$ and
    $\beta=5$.}   
    \label{fig:phen_polarized}
  \end{subfigure}
  \caption{\textbf{Temporal evolution of the normalized ratings for
      different values of $\boldsymbol{\alpha}$ and $\boldsymbol{\beta}$.}
    \subref{fig:phen_disorder} All $M$ normalized ratings $\hat{r}_{1i}(t)$ for user $u=1$ are shown.
    The configuration is disordered; all items are clicked on
    indifferently, and as a consequence all ratings fluctuate
    around the mean value $1/M$ (users are random clickers).
    \subref{fig:phen_cons_item} Again all $M$ normalized ratings
    $\hat{r}_{1i}(t)$ for user $u=1$ are shown.
    Since we are above the transition in $\beta$, the user tends to
    a single-item configuration, i.e., asymptotically only one item is
    clicked on; its normalized rating converges to $1$ while the others go
    to $0$. The comparison between panel (a) and (b) reveals the
    features of the~\textit{multiple--item to single--item} transition.
    \subref{fig:phen_cons_user}
    Here the ratings $\hat{r}_{u1}(t)$ of all
    users for the item $i=1$ are shown. This is a consensus configuration;
    all users, which asymptotically tend to a single-item state,
    click on the same item $i=1$.
    \subref{fig:phen_polarized} Again the ratings $\hat{r}_{u1}(t)$ of all $N$ users for item $i=1$
    are shown. Since $\alpha>\alpha_c$ users are polarized; some of them tend to click
    only on item $i=1$ but others tend to click only on a different item $i'$.
    The comparison between panel (c) and (d) reveals the
    features of the \textit{consensus-polarization} transition.}
  \label{fig:phenomenology}
\end{figure*}
First we focus, without lack of generality, on a
specific user, $u=1$, and we study the behavior of the ratings
$\hat{r}_{1i}(t)$ for such a user.
\ref{fig:phen_disorder} shows the temporal evolution of
these quantities for
$M=25$ items and $\alpha=\beta=0.5$: all normalized ratings converge
to the value $1/M$, meaning that the user under consideration equally
rates all possible opinions.
This corresponds to a disordered configuration in which the
algorithm provides random recommendations and the users behave as
random clickers.
The situation radically changes when the popularity bias is increased to
$\beta=5$, as shown in \ref{fig:phen_cons_item}.
In this case one of the normalized
ratings converges to one, while all the others go to zero, meaning
that the user ends up always choosing the same opinion.
Hence the symmetry among items,
found for small values of $\beta$, breaks down when the popularity bias is increased.
Analogously, we can focus on a given item (we choose the first without
lack of generality) and study the rating that different users give to it.
\ref{fig:phen_cons_user} shows the evolution of the $N$
normalized ratings $\hat{r}_{u1}(t)$, corresponding to the various
items, for $\alpha=0.5$ and $\beta =5$. We observe consensus among
users, since for all of them the normalized rating of item $1$
converges to one, while all other normalized ratings (not shown) go to zero.
Finally, we report in \ref{fig:phen_polarized} the
behavior of $\hat{r}_{u1}(t)$ when also the similarity bias $\alpha$
is set to 5.
The symmetry among users breaks down.
While for small $\alpha$ they all act the same, here their behavior is
heterogeneous: some of them maximally rate the first item, while for
others the normalized rating corresponding to such an opinion vanishes,
thus giving rise to a polarized configuration in which users are divided
into groups depending on the opinion they support.
\section{Phase diagram and associated transitions}
When the control parameters $\alpha$ and $\beta$ are varied between
the limit values discussed above, phase-transitions take place,
associated to distinct symmetry breakings. For small values of the
biases both users and items are completely symmetric and the system is
in a disordered phase where all users rate all items in the same
way. The popularity bias $\beta$ is responsible for the item-symmetry
breaking: when this parameter gets sufficiently large each user only
clicks on a specific item. Analogously, an increase of the similarity
bias $\alpha$ breaks the user-symmetry, leading to heterogeneous user
behavior. The polarized phase observed for large values of the biases
emerges when both symmetries are broken, while consensus occurs when
only the item-symmetry is broken. Note that no phase where only the
user-simmetry is broken is possible: if all items are equally rated by
each user, necessarily all users are similar. In this section we show
how these considerations can be made, by analytical and numerical
means, more grounded and precise.
\subsection{Master equation}
The master equation for the probability distribution of $r_{ui}$, $Q(r_{ui},t)$ is
\begin{equation}
    \dfrac{d}{dt}Q(r_{ui},t)=\frac{1}{\delta t} \left[
    \begin{multlined}[t]
        Q(r_{ui}-1,t)R_{ui}(r_{ui}-1)+\\
        -R_{ui}(r_{ui})Q(r_{ui},t),
    \end{multlined}
    \right]
        \label{eq:master_equation}
\end{equation}
from which the drift coefficient is readily obtained 
\begin{equation}
    \nu_{ui}=\frac{d\mean*{r_{ui}}}{dt}=\frac{1}{\delta t}\mean*{R_{ui}} = N \mean*{R_{ui}}.
        \label{eq:drift_generale}
\end{equation}
It follows the expression for the drift of the normalized ratings
\begin{equation}
    \hat{\nu}_{ui}=\frac{d\mean*{\hat{r}_{ui}}}{dt} \approx \frac{1}{t+Mr_0}\qua*{N\mean*{R_{ui}}-\mean*{\hat{r}_{ui}}}.
        \label{eq:drift_normalized_generale}
\end{equation}
Detailed computations of Eqs.~(\ref{eq:master_equation}-\ref{eq:drift_normalized_generale}) are
reported in Appendix~\ref{Master}.  Focusing on the long time behavior
of the system, diffusion can be neglected and the evolution of the
normalized ratings can be approximated by means of the drift terms only
\begin{align}
    \frac{d\hat{r}_{ui}}{dt}&\approx\frac{1}{t+Mr_0}\qua*{N R_{ui}-\hat{r}_{ui}}
        \label{eq:drift}
\end{align}
The stationary solutions of the dynamics are those for which the time
derivative reported above is equal to zero, that is
$NR_{ui}-\hat{r}_{ui}=0$. It is easy to show (see
Appendix~\ref{Stationary}) that Disorder, Consensus and Polarization
are solutions of this equation. In particular, these solutions are
defined, in terms of normalized ratings, as:
\begin{itemize}
    \item \textbf{Disorder}: all ratings are equal,
      $\hat{r}_{ui}=1/M$, $\forall u$, $\forall i$;
    \item \textbf{Consensus}: the ratings for one item are 1 for all
      users ($r_{ui}=1$, $\forall u$) while the ratings for all other
      items are 0 for all users ($r_{uj}=0$, $\forall u$, $\forall
      j\ne i$);
    \item \textbf{Polarization}: there are at least two groups of
      users, labeled by $k$ and $k'$, with consensus within each group
      ($r_{ui_k}=1$, $r_{uj}=0$, $\forall j\ne i_k$, $\forall u \in
      k$), but the selected item is different for different groups
      ($i_k \ne i_{k'}$). It is important to remark that in
      the polarized phase, the interaction induced by the
      recommendation algorithm is such that users sharing
      a different opinion do not influence each other.
\end{itemize}

As shown in Appendix~\ref{Stationary}, also other
configurations, where $0<m<M$ items are equally rated, have a null
drift. In order to understand which are the true stationary states of
the system it is necessary to consider their stability.
\subsection{Stability analysis}
\subsubsection{The transition from multiple--item to single--item}
Let us consider the case $\alpha=\infty$. From the phenomenological
considerations presented above we expect the disordered solution
occurring for $\beta=0$ to be stable also for small $\beta$, while
users should stick to a single item for larger values of the
popularity bias. In the disordered phase all items are equally likely
to be clicked on. A disordered (multiple--item) configuration is then
described by all normalized ratings equal to $1/M$ with small
fluctuations $\epsilon_{ui}$:
\begin{flalign}
    \forall (u,i) && \hat{r}_{ui}=\dfrac{1}{M} + \epsilon_{ui}. && |\epsilon_{ui}| \ll \frac{1}{M}
        \label{eq:disorder_sol}
\end{flalign}

The fluctuations can be either positive or negative, and they are constrained
by the normalization of ratings
\begin{equation}
  \sum_i \hat{r}_{uj} = \sum_i \left( \dfrac{1}{M} + \epsilon_{uj} \right)=1 \implies \sum_j \epsilon_{uj}=0.
  \label{sumrule}
\end{equation}
By plugging~\eqref{eq:disorder_sol} into
\eqref{eq:drift}, recalling that
$r_{ui}/\sum r_{ui}=\hat{r}_{ui}/\sum\hat{r}_{ui}$ and expanding for small
$\epsilon_{ui}$ it is easy to show (see Appendix~\ref{DisorderSingle})
that
\begin{equation}
    \frac{d\epsilon_{ui}}{dt} \propto (\beta-1) \epsilon_{ui}.
        \label{stab}
\end{equation}
If $\beta<1$, fluctuations $\epsilon_{ui}$ are exponentially
suppressed. The system is then always driven back to the disordered
solution, which is therefore stable. When $\beta>1$ fluctuations are
instead amplified and the multiple--item solution is unstable. By a
similar argument (see Appendix~\ref{DisorderSingle}) it is possible to
show that the single-item solution, for which in the large time limit
$\hat{r}_{ui}=1$ and $\hat{r}_{uj}=0\ \forall \ j\neq i$, is
stable when $\beta>1$, while it is unstable if $\beta<1$. Moreover,
solutions characterized by users equally rating more than 1 but less
than $M$ items are found to be unstable both for $\beta<1$ and
$\beta>1$, showing the nonexistence of a fourth stable phase. We thus
conclude that at $\beta_c = 1$ a transition between the multiple--item
and the single-item solution occurs, associated to the breaking of
symmetry among items. In Appendix~\ref{DisorderSingle} we show that
the same picture applies in the general case $\alpha<\infty$.
\subsubsection{The transition from consensus to polarization}
As for the transition in $\beta$, we can study the transition in
$\alpha$ by looking at the stability of the consensus and polarization
solutions. Since we already know that for $\beta<1$ the single-item
solution is never observed, we can assume $\beta>1$; for simplicity
here we set $\beta=\infty$, while we refer to
Appendix~\ref{ConsensusPolarization} for the general case. We assume
users close to the single-item solution, i.e.,
\begin{equation}
    \begin{cases}
        \hat{r}_{ui}=1-\epsilon & \\
        \hat{r}_{uj}=\dfrac{\epsilon}{M-1} &~~~ \forall j \ne i,
    \end{cases}
        \label{eq:singleitemsol}
\end{equation}
with $0<\epsilon \ll 1$. Note that here we made the assumption that
$\epsilon$ does not depend on $u$ and $i$, but the same results can be
obtained also considering the more general case of fluctuations of the
form $\epsilon_{ui}$.\par

In the consensus configuration all users are aligned along the same
item $i$ and so all similarities are approximately equal to 1. As a
consequence, since $\beta=\infty$, we can write \eqref{eq:drift} as
\begin{align*}
    \frac{d\hat{r}_{ui}}{dt}&\approx \dfrac{1}{t+Mr_0} \qua*{\dfrac{\sum_v s_{uv}^{\alpha}\delta_{i,i_v} }{\sum_w s_{uw}^{\alpha}}-\hat{r}_{ui}}=\\
    &=\frac{1}{t+Mr_0}\qua*{1-(1-\epsilon)}=\frac{\epsilon}{t+M r_0}>0,
\end{align*}
where we used the fact that all similarities are equal to 1 and
$i_v=i$ for all users. Since this quantity is always positive
regardless of $\alpha$, the normalized rating along $i$ keeps
increasing, asymptotically converging to the consensus solution
$\hat{r}_{ui}=1$. Considering instead an item $j \neq i$, we get for
the drift
\[
    \frac{d\hat{r}_{uj}}{dt}=\frac{1}{t+Mr_0}\qua*{0-\frac{\epsilon}{M-1}}<0,
\]
meaning that the normalized ratings of the other items consistently go
to zero. These results imply that the consensus solution is an
attractor of the dynamics for any value of the similarity bias
$\alpha$.\par

We can then turn to the study of the polarized state. In this case the
similarity between two users polarized on the same item, is again 1,
while for users polarized on different items it holds
\begin{flalign*}
    && s_{uv}=\dfrac{2(M-1)\epsilon-M
    \epsilon^2}{1-2\epsilon+M\epsilon^2}=O(\epsilon) && \text{if~} i_u \neq i_v,
\end{flalign*}
where $i_u$ is the item user $u$ is polarized on. Assuming for
simplicity $K<N/2$ distinct users polarized on item $i_u$, while the
remaining $N-K$ polarized along $i_v$, \eqref{eq:drift} reads
\[
    \frac{d\hat{r}_{u i_u}}{dt} \approx\dfrac{1}{t+Mr_0} \left[ \dfrac{K
  }{K+(N-K) \epsilon^{\alpha}}-(1-\epsilon) \right].
\]
This quantity is negative for $\alpha<1$, while it is positive for
$\alpha>1$. Analogously, the drift of the rating of user $u$ along
item $i_v$ is
\[
\frac{d\hat{r}_{u i_v}}{dt} \approx\dfrac{1}{t+Mr_0} \left[ \dfrac{K\epsilon^{\alpha} }{(N-K)+K\epsilon^{\alpha}}-\dfrac{\epsilon}{M-1} \right].
\]
which is positive for $\alpha<1$ and negative for $\alpha>1$. These
results imply that for $\alpha<\alpha_c=1$ the $K$ users polarized
along $i_u$ will not remain polarized as the system evolves, finally
polarizing along the item $i_v$ shared by the majority of
agents. Conversely, for $\alpha>\alpha_c$ the drift reinforces
the minority and a polarized state consisting of two different group
of users emerges. The analysis thus indicates that for $\alpha<1$ only
the consensus state is stable, while we expect to observe both
consensus and polarization above the critical value.\par

The overall phase--diagram of the model is summarized in \ref{fig:stream}.
\subsubsection{The simplest case: \texorpdfstring{$N=M=2$}{N=M=2}}
\begin{figure*}[t]
    \centering
    \includegraphics[width=\textwidth]{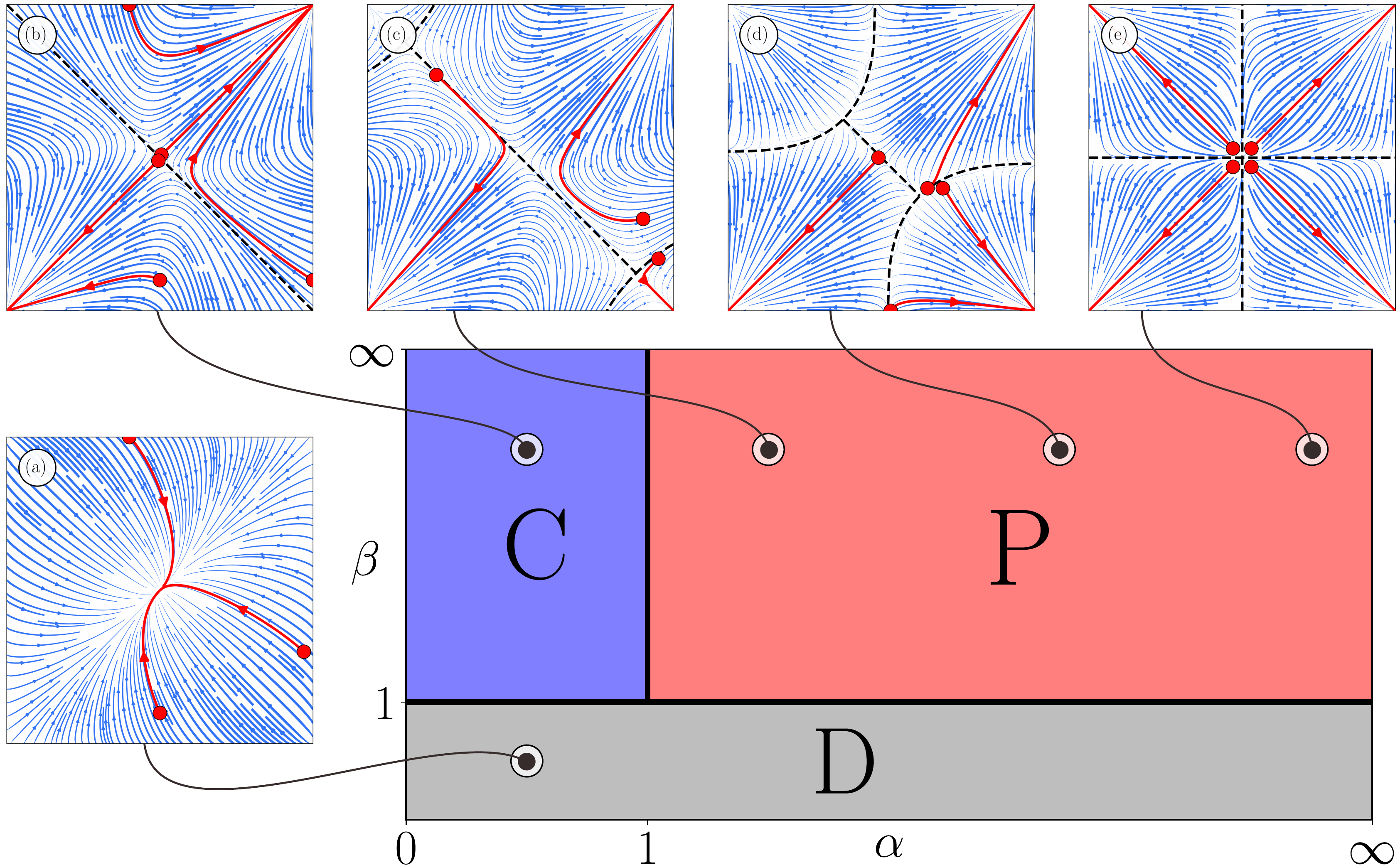}
    \caption[Phase Diagram of the model.]
            {\textbf{Phase diagram of the model.}
              Graphical representation of the phase diagram. The
      $(\alpha,\beta)$ plane is split up into three regions according
      to the distinct observed phases. The $\beta=1$ line divides the
      multiple--item (Disorder) regime from the single--item one. This
      is in turn split up into the Consensus phase, where all users
      agree on the same item, and Polarization phase where users stick
      to different items. The line at $\alpha=1$ is exact only in the
      $N\to\infty$ limit, while for finite size systems the transition
      (crossover) is observed for larger \(\alpha\).
      The panels around the diagram
      show the ratings streamlines for $N=2, M=2$ at different
      $(\alpha,\beta)$ values. The red streamlines represent the
      evolution of the ratings when starting from the respective red
      dots, showing the fixed points of the dynamics. The
      basins of attractions are separated by black dashed lines.
      In the Disorder phase the only fixed point is the point
      $(\hat{r}_{1, 1}, \hat{r}_{2, 1}) = (0.5, 0.5)$, as in panel (a).
      In the Consensus phase
      $(\hat{r}_{1, 1}, \hat{r}_{2, 1}) \in \{(0, 0), (1, 1)\}$,
      as in panel (b). Above
      $(\alpha, \beta) = (1, 1)$ two new attraction basins arise which
      are related to the emergence of Polarization states,
      $(\hat{r}_{1, 1}, \hat{r}_{2, 1}) \in \{(1, 0), (0, 1)\}$, as in panel
      (c)--(d). Eventually, when $\alpha \to \infty$ (e) the phase
      space is equally divided into Consensus and Polarization.}
            \label{fig:stream}
\end{figure*}
We can easily visualize the transitions in the simplest possible case
$N=M=2$. In this situation the state of the system is described in
terms of two variables only. Indeed from the normalization of the
ratings it follows that we can focus on the ratings users give to the
first item, i.e., $\hat{r}_{11}$ and $\hat{r}_{21}$. This allows to
visualize the drift given by \eqref{eq:drift} through a stream plot in
the square ($0 \le \hat{r}_{11} \le 1$, $0 \le \hat{r}_{21}\le
1$). The disordered state D corresponds to the point
($\hat{r}_{11}=\hat{r}_{21}=0.5$), consensus C to the two points
($\hat{r}_{11}=\hat{r}_{21}=1$, $\hat{r}_{11}=\hat{r}_{21}=0$), while
polarization P corresponds to the points ($\hat{r}_{11}=1$,
$\hat{r}_{21}=0$) and ($\hat{r}_{11}=0$,
$\hat{r}_{21}=1$). \ref{fig:stream}a shows the vector field
$\vec{\hat{\nu}}=(\hat{\nu}_{11}, \hat{\nu}_{21})$ for $\alpha=0.5$
and $\beta=0.5$. As discussed above, all stream lines point toward the
disordered state D, which is the only attractor of the dynamics. This
is also shown by three different trajectories (in red) which all end
up in the central point of the stream plot. The situation changes by
increasing the popularity bias above the critical value $\beta_c=1$ to
$\beta=1.5$, as shown in \ref{fig:stream}b. Disorder stops to be an
attractor and all the stream lines point toward the two consensus
configurations C. The transition is abrupt, since as soon as the
popularity bias exceeds $\beta_c$, the attraction basin of the
disordered state disappears, while that of consensus occupies all the
phase space.\par

In the same way we can investigate the transition driven by the
similarity bias. Starting from \ref{fig:stream}b, where consensus is
the only attractor, we increase the similarity bias to $\alpha=1.5$
(\ref{fig:stream}c), a value above the critical value
$\alpha_c=1$. This makes the polarized state P emerge, although its
basin of attraction still remains small compared to that of
consensus. For larger values of $\alpha$ the basin of attraction of
polarization grows (\ref{fig:stream}d), reaching the same size of the
attraction basin of Consensus in the limit of infinite similarity bias
(\ref{fig:stream}e): the phase space splits into four quadrants, two
belonging to the Consensus attractor, the other two to the
Polarization one. Starting from a fully disordered initial condition
(the center of the square) and neglecting diffusion, in the
$\alpha=\infty$ limit we expect to reach consensus in half of the
realizations of the process. For smaller values of $\alpha>1$ this
probability will be larger than $1/2$.
\subsection{Numerical investigation of the phase transitions}
The phase--diagram deduced in the previous subsections and sketched in
\ref{fig:stream} can be validated by means of numerical simulations of
the model behavior. In particular, it is possible to define two order
parameters, related to the variance of normalized ratings, whose
values mark the two transitions, associated to the breakings of user
or item symmetries.  See Appendix~\ref{OrderParameters} for
details. While the transition controlled by $\beta$ occurs as
expected, with an abrupt jump of the order parameter around $\beta=1$,
the analysis of the consensus-to-polarization transition, controlled
by $\alpha$, requires more care, since above $\alpha_c=1$, both
consensus and polarization solutions are stable.\par

In the $N=M=2$ case, in the limit of large $\alpha$, the phase space
splits into four equally-sized regions, two belonging to the basin of
attraction of consensus, the others to the basin of attraction of
polarization. The initial condition we adopt, with all ratings equal,
lies in the center of the phase space, corresponding to a perfectly
disordered configuration.  This means that, neglecting diffusion, the
very first random click completely determines whether the system
evolves toward consensus or polarization. Since the first click
corresponds, in the stream plot, to a step along one of the diagonals
with equal probability, we expect consensus to be reached with
probability $P_C=1/2$.\par

In the case of generic $N$ and $M$ and $\alpha=\infty$ the picture is
similar. Since every user is completely independent from the others
he/she gets polarized along one item uniformly selected at random
among the $M$ possible values. Global consensus is reached only if, by
chance, the selected item is the same for all users. All other
configurations correspond to polarization. The probability $P_C$ that
all users randomly choose the same item is
\begin{equation}
    P_C=\frac{M}{M^N}=M^{-N+1}.
        \label{eq:prob_consensus}
\end{equation}
In the limit of large $M$ (or large $N$) this probability of consensus
$P_C$ vanishes: Polarization is the only stationary state actually
reached by the dynamics. Consensus gets harder and harder to be
observed due to entropic effects. This can be also seen by noticing that the consensus corners form a countable set, while the polarization corners are an uncountable one, since the latter can be also seen as the set of all infinite binary sequences.\par

\eqref{eq:prob_consensus} implies that, exactly as for $N=M=2$, for
generic $N$ and $M$ and $\alpha=\infty$ the phase-space is divided in
$M^N$ quadrants of which only $M$ lead to consensus. By analogy with
the $N=M=2$ case we expect that, also in the generic case, the basin
of attraction of polarization starts gradually growing from the
``corners'' of the phase-space as $\alpha$ becomes larger than 1,
eventually invading the whole quadrants. Note that the system is
always initialized in the center of the phase space, because all
ratings are initially equal. Then, the first random click moves it
toward the periphery of the phase space, where the polarization basin
lies. Since in the large system limit, there are infinitely more
polarized corners than consensus ones, we expect random fluctuations
to lead the system to a polarized state very easily and thus the
consensus probability to go to zero very rapidly also for values of
$\alpha$ close to the critical point $\alpha_c=1$. The scenario just
described is confirmed by numerical simulations.
\begin{figure*}
    \centering
    \begin{subfigure}[b]{0.475\textwidth}
        \centering
        \includegraphics[width=\textwidth]{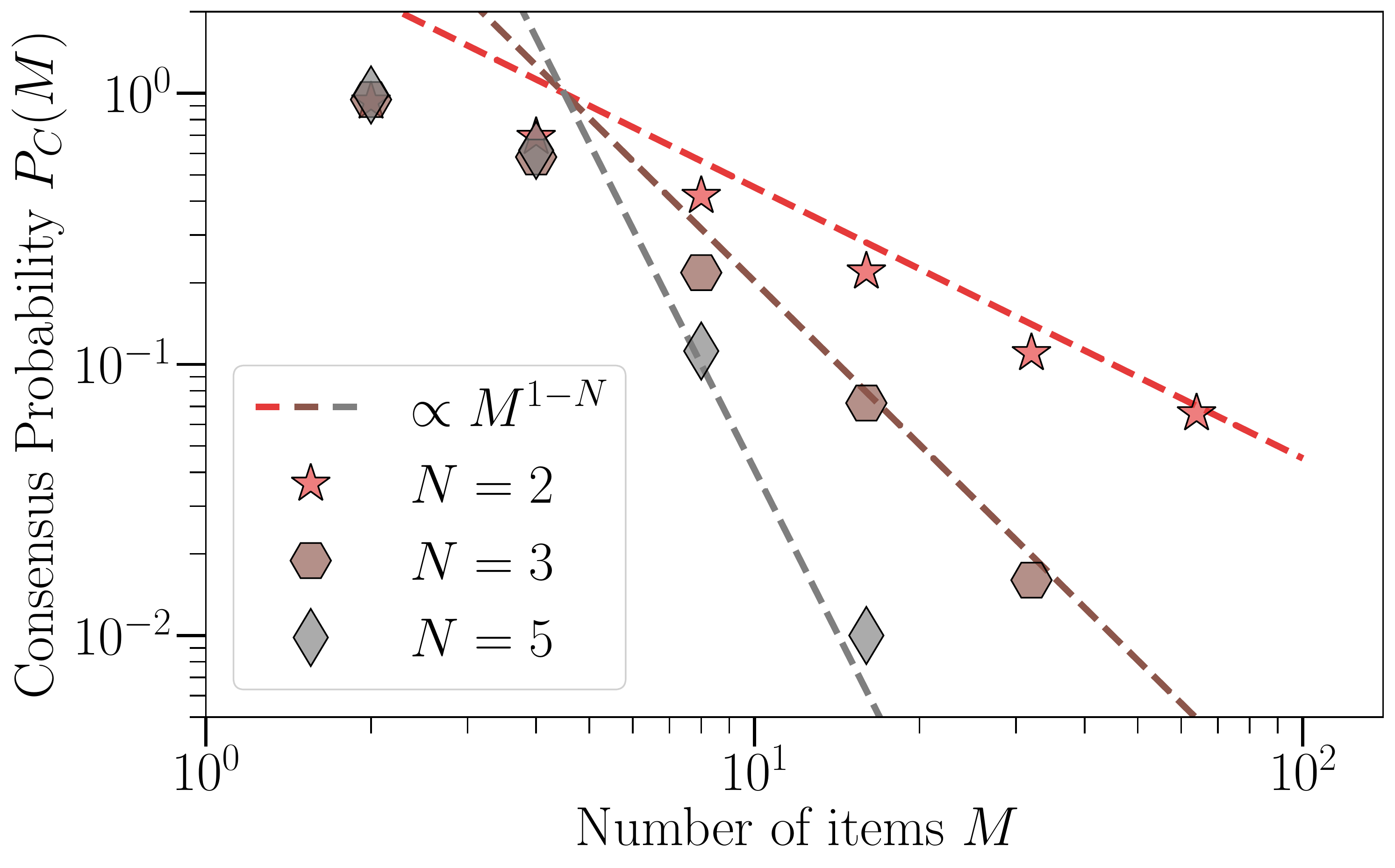}
        \caption{}\label{fig:Pc(N)}
    \end{subfigure}
    \hfill
    \begin{subfigure}[b]{0.475\textwidth}  
        \centering 
        \includegraphics[width=\textwidth]{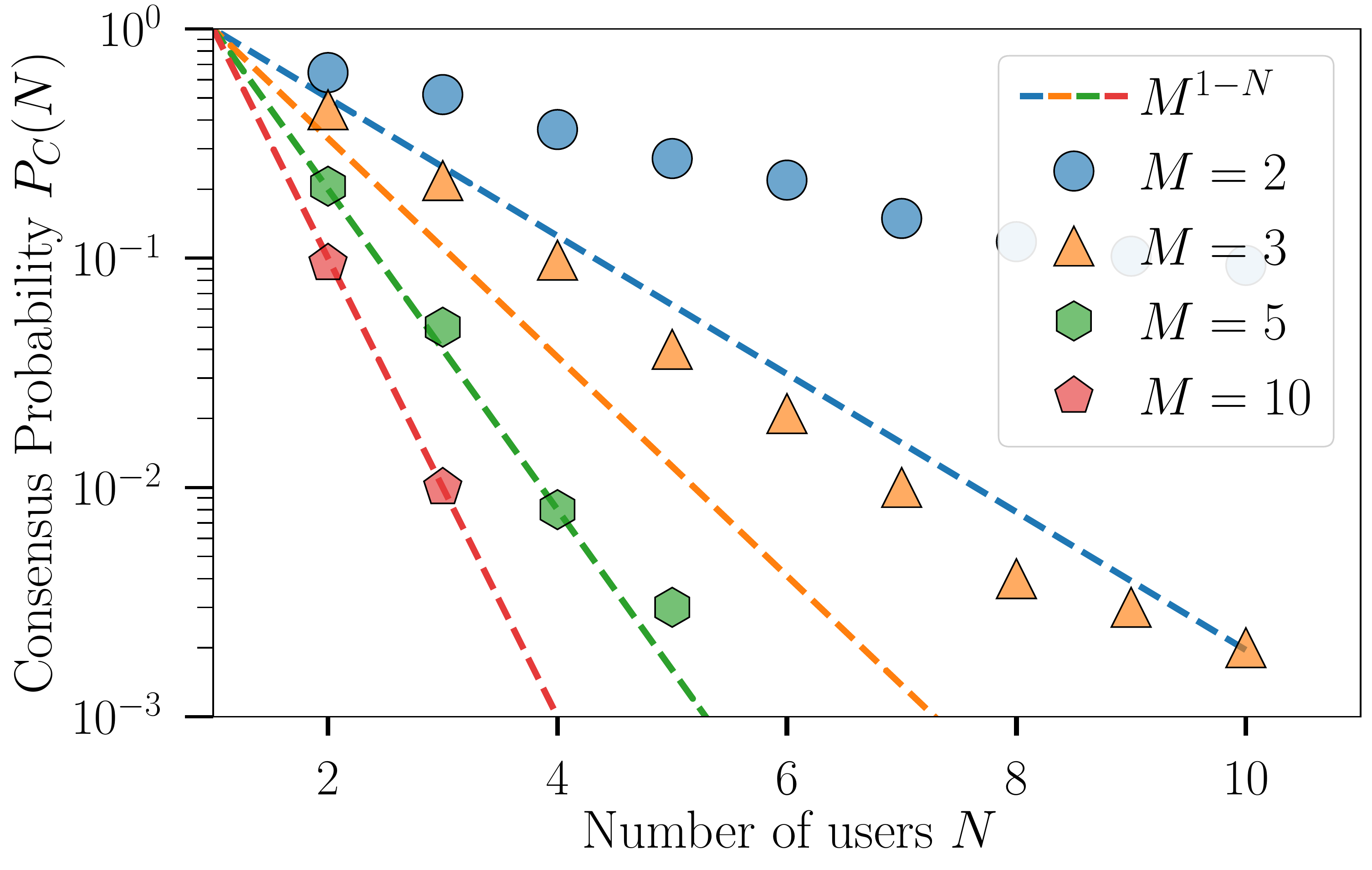}\\
        \caption{}\label{fig:Pc(M)}
    \end{subfigure}
    \caption{\textbf{Probability of consensus as a function of $M$ and
        $N$.} Fraction of 1000 runs going to
      Consensus. Panel \subref{fig:Pc(N)} shows that, for fixed
      $\alpha=10$, $P_C$ tends to 0 as $N$ is increased. Larger values
      of $M$ imply a better agreement with \eqref{eq:prob_consensus},
      derived for $\alpha=\infty$. Panel \subref{fig:Pc(M)} shows
      that, already for $\alpha=2$, the power--law dependence of $P_C$
      on $M$ with exponent $1-N$ (see \eqref{eq:prob_consensus}) is
      obeyed for sufficiently large $M$, with an $N$-dependent
      prefactor.}\label{fig:scaling_Pc}
\end{figure*}
~\ref{fig:Pc(N)} and~\ref{fig:Pc(M)} show a comparison between
\eqref{eq:prob_consensus} and the probability of consensus measured in
numerical simulations. For fixed and large enough $M$, the exponential
decay in $N$, $M^{-N+1}$, is perfectly recovered in simulations, while
for small $M$ we observe strong discrepancies. This behavior
derives from the fact that, as discussed in
Appendix~\ref{OrderParameters}, by increasing $N$ only, the transition
gets sharper, but occurs at values of $\alpha$ larger than 1.
Conversely, when $M$ is increased, the transition moves toward
$\alpha_c=1$ while also getting sharper. Keeping instead
$N$ fixed and looking at the consensus probability as a function of
$M$ we observe a good agreement for what concerns the scaling
exponent, $1-N$, although with a much larger prefactor. In any case,
we can conclude that, for $N$ and $M$ sufficiently large, entropic
effects make consensus very hard to be reached as soon as
$\alpha>\alpha_c=1$. Hence in large systems the phase diagram is
composed of three pure phases: disorder, consensus and polarization.
\section{Critical recommendations}
The analytical approach and the numerical simulations show that the
collaborative--filtering model is characterized by three distinct
phases: Disorder, Consensus and Polarization. Only in the latter the
algorithm really provides personalized recommendations. Indeed, in the
disordered phase users get completely random recommendations, while in
the consensus phase there is no personalization, as all users receive
exactly the same suggestion. Conversely, in the polarization regime
users spontaneously split into groups, each characterized by a
different recommended item. Thus, in this phase the
algorithm provides to each user personalized recommendations
perfectly in line with his/her past choices. Note, however, that
in the long run each user is exposed only to a single item.
Users are
trapped into a filter bubble preventing them from being exposed to all
other items.\par

Understanding if and how personalized recommendations can be obtained
without users being completely stuck in a filter bubble is of crucial
importance. With this goal we focus on the transition between multiple--item
and single--item, where an intermediate behavior is expected to be
found.
This transition corresponds to the line $\beta=\beta_c=1$. For
such value of the popularity bias, the probability 
for user $u$ to click on item $i$ reads, from \eqref{eq:transition_prob},
\begin{align*}
    R_{ui} &= \frac{1}{N}\sum_{v=1}^N\frac{s_{uv}^{\alpha}}{\sum_ws_{uw}^{\alpha}}\frac{r_{vi}}{\sum_j^M r_{vj}}=\\
    &\approx \frac{1}{N}\sum_{v=1}^N\frac{s_{uv}^{\alpha}}{\sum_ws_{uw}^{\alpha}}\frac{r_{vi}}{t+Mr_0}.
\end{align*}

\begin{figure*}
    \centering
    \begin{subfigure}[t]{0.475\textwidth}
        \centering
        \includegraphics[width=\textwidth]{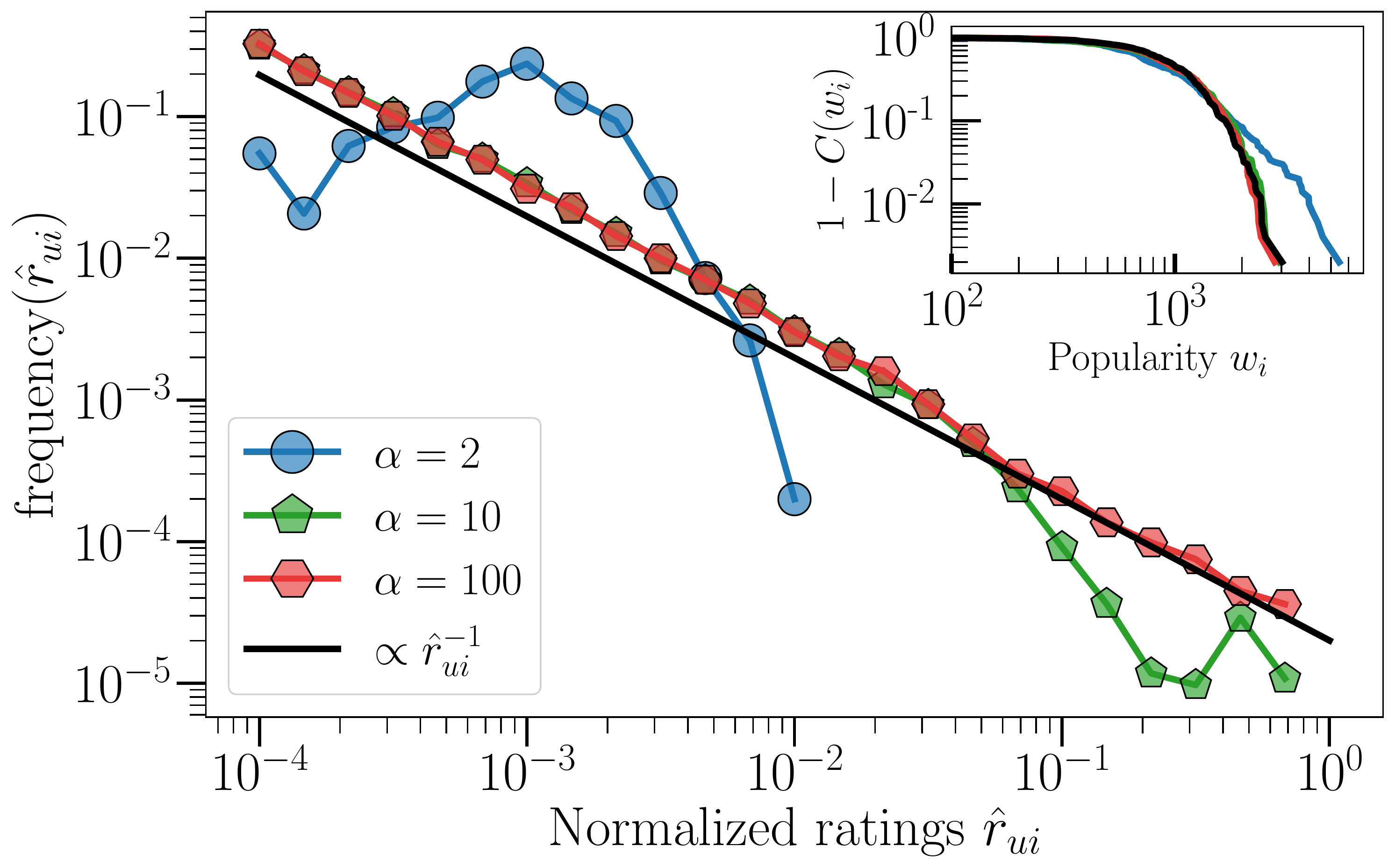}
        \caption{}
        \label{fig:beta1_alpha_large}
    \end{subfigure}
    \hfill
    \begin{subfigure}[t]{0.475\textwidth}
        \centering
        \includegraphics[width=\textwidth]{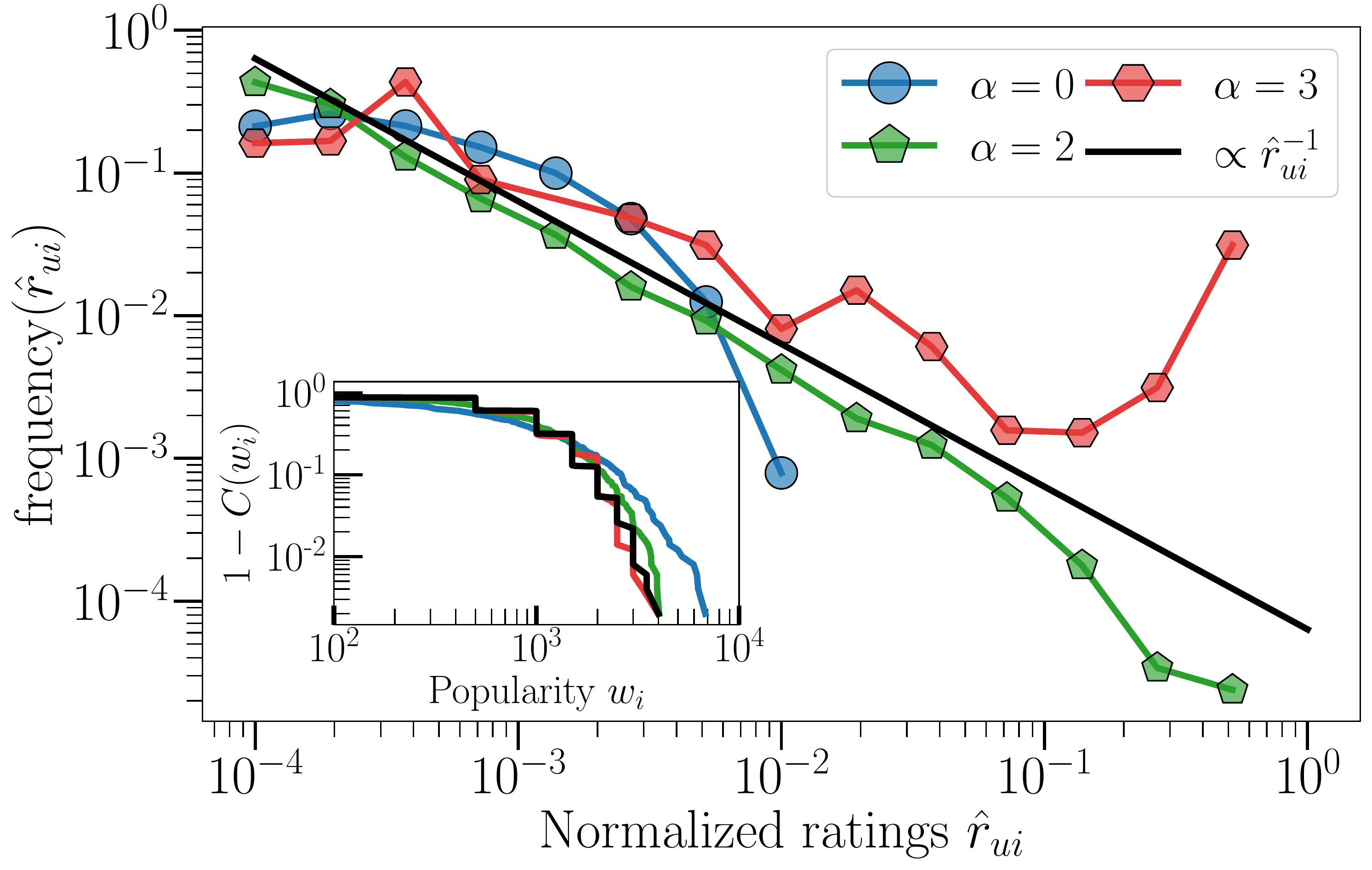}
        \caption{}
        \label{fig:beta1_alpha_small}
    \end{subfigure}
  \caption{\textbf{Distributions of the normalized ratings for
      $\beta=1$}. Left panel, (a): distribution of normalized ratings for
    $N=1000$, $M=500$, $\beta=1$ and different values of $\alpha$.
    The system is initialized with initial conditions
    $r_0^{(1)}=1/(M-1)$. In this case we expect a power-law
    distribution with exponent $-1$ for large values of $\alpha$
    (straight line), which we observe already for $\alpha=10$. When
    $\alpha=2$, the approximation of independent users breaks and the
    distribution is no more approximated by a power law. The inset
    shows the complementary cumulative distribution of the
    popularity. For large $\alpha$ it coincides with that of a Polya
    urn (black line) since users are independent, while as $\alpha$
    decreases it becomes broader, due to the emergence of correlations
    among users.
    Right panel, (b): as in the left panel, but initializing
    the system with initial conditions $r_0^{(2)}=1/N(M-1)$. In this
    case the model approximately follows a Polya Urn dynamics for
    $\alpha=0$; as expected, the distribution is close to a power law
    with exponent $-1$ (straight line). For larger values of $\alpha$,
    when users are mostly independent, a peak at
    $\hat{r_{ui}}\approx1$ appears in the power law distribution. Also
    in this case, the inset shows the complementary cumulative
    distribution of the popularity, that becomes more and more similar
    to that obtained from independent Polya Urns (black line) as
    $\alpha$ increases.}
  \label{fig:power_law}
\end{figure*}

In the disordered phase all normalized ratings $\hat{r}_{ui}$ for a
given user $u$ are concentrated around the value $1/M$ and, for large
$t$, they are described by a delta distribution. In the polarized
phase they are instead described by the superposition of two delta
functions, one in zero and the other in one, corresponding to the
winning opinion. We want to understand if, on the critical line, the
distribution of ratings assumes a nontrivial form between these two
limits.
We first consider the limit $\alpha\to\infty$,
where each agent is coupled only to his/her past,
different agents being completely independent. We can focus on just
one user and set $N=1$. In this way the transition rates become
\begin{equation}
    R_{ui}=R_i = \frac{r_i}{t+Mr_0}
        \label{eq:transition_polya}
\end{equation}
where we indicate for simplicity with $r_i$ the ratings of the user
under consideration. \eqref{eq:transition_polya} can be seen as the
transition rate for a Polya Urn model with balls of $M$ distinct
colors \cite{polya1930quelques}. Indeed, we can interpret $r_i$ as the
number of balls of color $i$ inside the urn and $R_i$ is the
probability of randomly extracting a ball of such a
color~\footnote{Since $r_0$ is noninteger, the interpretation in terms
  of balls in an urn does not strictly apply here. However, the theory
  for Polya Urns works also for real $r_i$.}.
Since at each time step exactly one item is clicked on and
the corresponding rating is increased by a unit, there is a perfect
mapping between the collaborative--filtering model for $\alpha=\infty$
and $\beta=1$ and a Polya urn with reinforcement parameter $S=1$. This
implies that the probability $P(\vec{\hat{r}})$ of observing a
normalized rating vector $\vec{\hat{r}}=(\hat{r}_1,\ldots,\hat{r}_M)$
is given by
\begin{equation}
    P(\vec{\hat{r}})=\frac{\prod_i^M r_i^{{\vec{r_0}}/{S}-1}}{\mathcal{D}\ton*{{\vec{r_0}}/{S}}},
        \label{Polya}
\end{equation}
where $S=1$ is the reinforcement parameter of the Polya Urn,
$\vec{r_0}=(r_0,\ldots,r_0)$ is the vector of initial conditions and
$\mathcal{D}(\cdot)$ is the Multivariate Beta
Function~\cite{polya1930quelques,Mauldon1959}.
\par
The distribution of a single normalized rating $P(\hat{r}_i)$ can be
obtained by marginalizing \eqref{Polya} over the remaining $M-1$
ratings, obtaining
\begin{align*}
    P(\hat{r}_i) =\frac{\hat{r}_i^{r_0-1}\ton*{1-\hat{r}_i}^{(M-1)r_0-1}}{B(r_0, (M-1)r_0)},
        \label{Polya}
\end{align*}
where $B(x,y)$ is the Euler Beta function. Depending on the value of
$r_0$ this distribution has different shapes (see
Appendix~\ref{BetaDirichlet} for details).
In particular, for $r_0=1/(M-1)$, it is a pure power--law with
exponent $-(M-2)/(M-1)$. This means that on the critical line
$\beta=1$, for $\alpha=\infty$ users are neither completely polarized
nor behaving randomly. Rather they show a non--trivial distribution of
the ratings, thus conciliating personalized recommendations with the
exploration of the whole item space. \ref{fig:beta1_alpha_large}
confirms this prediction also for values of the similarity bias
smaller than infinity: for $\alpha \approx 10$
users behave as if they were independent. We can see this
also by looking at the popularity $w_i=\sum_u r_{ui}$, whose
distribution quantifies how much different users agree on the same
items. When users are practically independent, we expect such a
distribution to be peaked around its mean value.
Indeed different users select different items as their favorite
and thus summing on all users gives, for all items, approximately the
same popularity value. This behavior is shown in the inset
of ~\ref{fig:beta1_alpha_large}, where we reported the
complementary cumulative distribution of the popularity in the large
$\alpha$ regime. \par

What happens when users cannot be considered effectively independent?
For $\alpha=0$ an approximate mapping to a Polya Urn
(see Appendix~\ref{BetaDirichlet}) yields a distribution of normalized
ratings perfectly analogous to \eqref{Polya}, with only $r_0$ replaced
by $r_0N$. As a consequence, $P(\hat{r}_i)$ is a Beta Distribution
(see Appendix~\ref{BetaDirichlet}), which
for initial condition $r_0=1/(N(M-1))$, decays as a
power--law $\hat{r}_i^{-1}$.
This behavior is
checked in ~\ref{fig:beta1_alpha_small}, where we show the
distribution of the ratings for various values of $\alpha$ and
$r_0=1/[N(M-1)]$. For $\alpha=0$ we observe deviations from the
predicted behavior, but as $\alpha$ is increased, the system more closely
follows a power law distribution.
Also in this case we show in the inset the complementary cumulative
distribution of the popularity,
which is broader than the one of a Polya urn for small $\alpha$, while it
becomes more and more similar to it as $\alpha$ increases.

In conclusion, even if for intermediate values of $\alpha$ an exact
mapping to a Polya Urn is not possible, what we observe numerically is
that the behavior of the system for $\beta=1$ and generic $\alpha$
is well described either by the $\alpha=0$ limit or the $\alpha=\infty$.
In all cases we find broad rating distributions.
Note that the behavior we observe on the critical line
  depends on the point where we cross it, i.e. if we move from
  disorder to consensus or from disorder to polarization. An analogous
  power law scaling is found only using a different $r_0$ in the two
  cases. In particular, when $\alpha$ is small, users shows a stronger
  degree of collective consensus, while for larger $\alpha$ they
  behave more individualistically.It is also
  important to remark that when $M, N$ are large enough, the
  transition occurs at $\beta_c=1$ independently of the number of
  users or items. This implies that the critical recommendation regime
  is stable when new users or items enter the system, a crucial
  requirement for applying the recommendation algorithm in realistic
  scenarios. Moreover, the standard implementation of the user-user
  collaborative filtering, corresponding to $\alpha=\beta=1$, lies on
  the critical line and it is thus in the optimal region of the
  algorithm's phase space.

%
    \section{Modeling music recommendations}
\begin{figure*}
    \centering
    \begin{subfigure}[b]{0.32\textwidth}  
      \includegraphics[width=\textwidth]{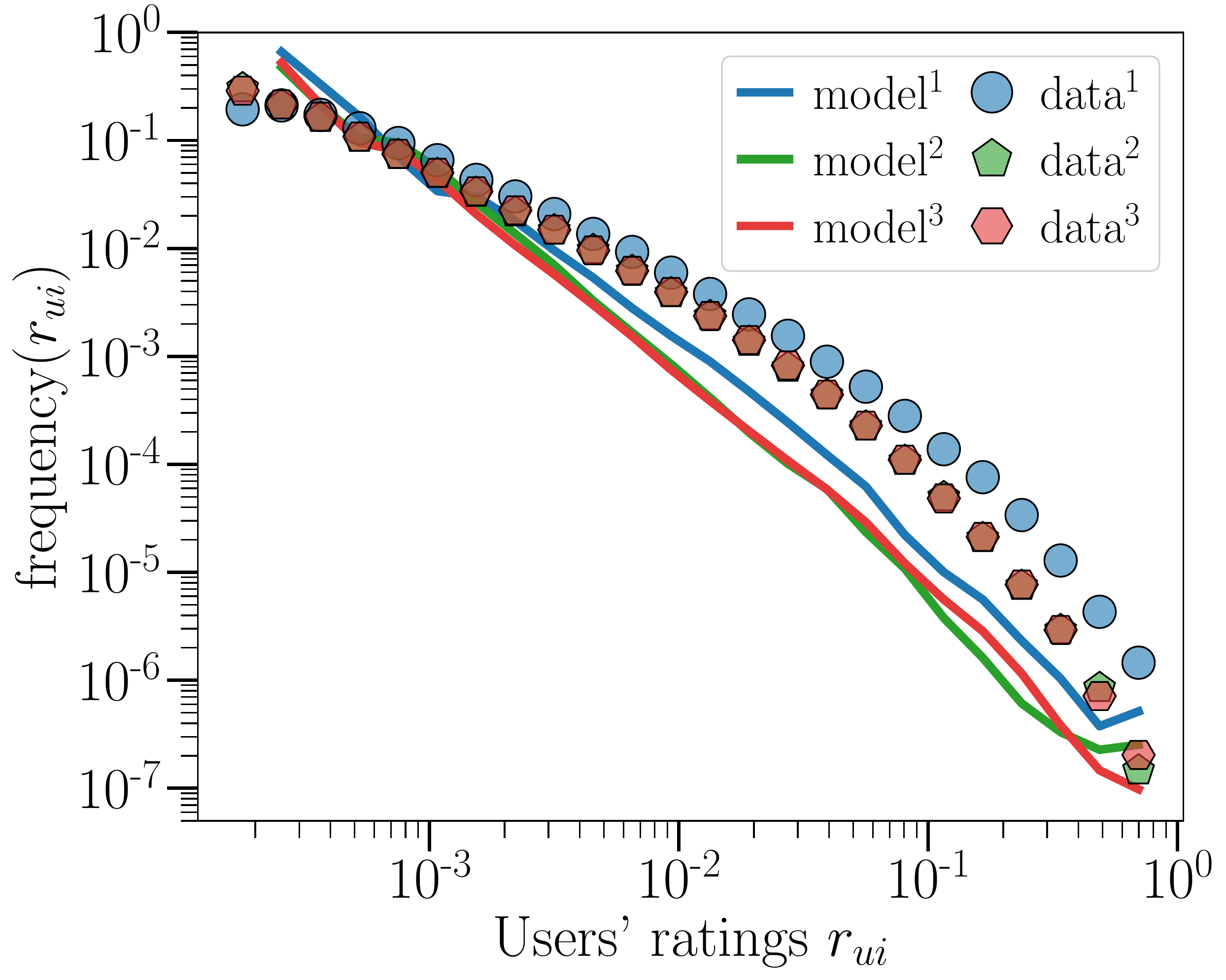}
      \caption{Ratings distributions.}
      \label{sfig:rating}
    \end{subfigure}
    \begin{subfigure}[b]{0.32\textwidth}  
        \includegraphics[width=\textwidth]{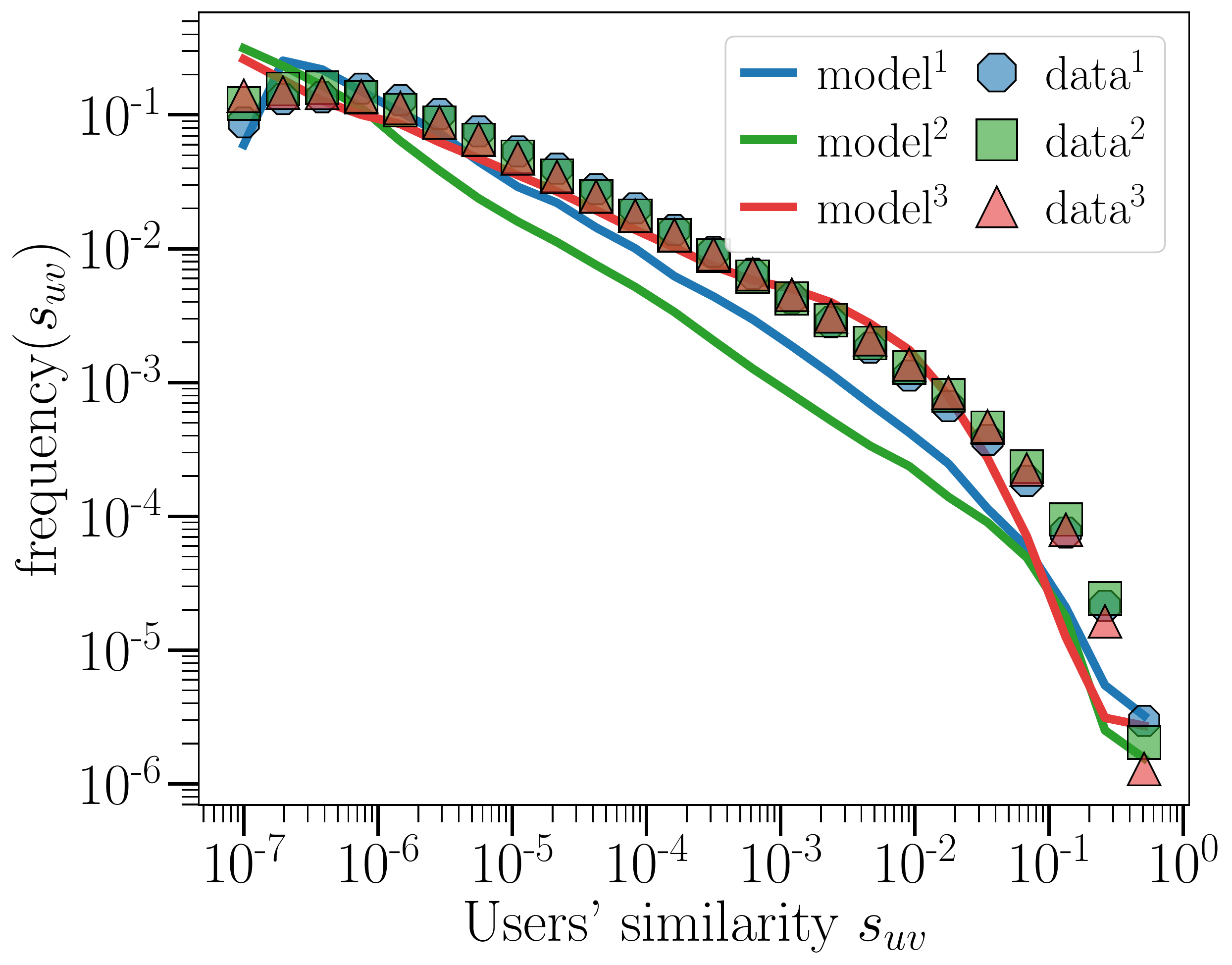}
        \caption{Similarities distributions.}
        \label{sfig:sim}
    \end{subfigure}
    \begin{subfigure}[b]{0.32\textwidth}  
        \includegraphics[width=\textwidth]{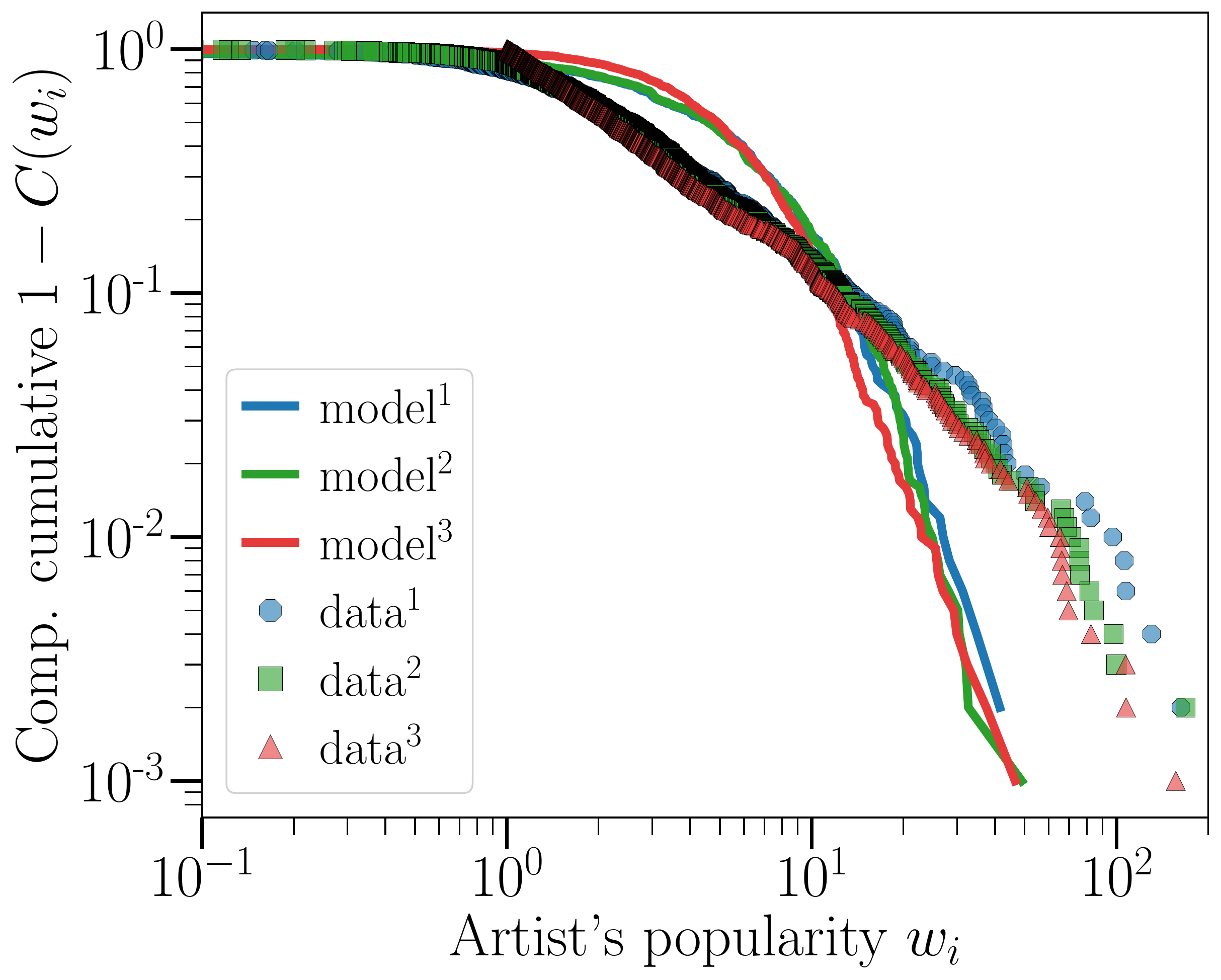}
        \caption{Popularities distributions.}
        \label{sfig:pop}
    \end{subfigure}
    \caption[Empirical data from last.fm]{\textbf{Empirical data from
        last.fm} \subref{sfig:rating}: Histogram of the empirical
      ratings from last.fm (symbols) for three combinations of $N,M$:
      $(N, M)^1=(1000, 500)$, $(N, M)^2=(2000, 1000)$, $(N,
      M)^3=(5000, 1000)$. Each rating represents how many times a given user
      listened to a particular artist. Distributions obtained by
      simulating the model with a fixed parameter $\alpha$:
      $\alpha^{\rm fix,1} = 1.74$, $\alpha^{\rm fix,2} = 1.78$,
      $\alpha^{\rm fix,3} = 1.98$ are shown as solid
      lines. \subref{sfig:sim}: Histogram of the cosine similarity for
      the same set of parameters of the previous figure. Empirical
      data are plotted as symbols; results from simulations of the
      model are shown as solid lines. \subref{sfig:pop}: Complementary
      Cumulative distribution of the popularity defined as
      $w_i=\sum_{u} r_{ui}$. Empirical data show a wide distribution
      of the popularity, a feature which cannot be fully recovered by our model.}.
    \label{fig:fit_lastfm}
\end{figure*}

The model we consider describes a
recommendation algorithm with implicit feedback, where the ratings
are computed from users' behavior and not directly from their
votes. This is the typical situation in online music platforms; in
such a context the number of times a user plays a song, i.e. what we
call rating, is a proxy of how much the user likes that song.
Thus, it is very natural to compare the model behavior with data
coming from an online music platform. The popular website last.fm
is a suitable platform for such a task, as it provides full listening
histories of a large amount of its users. These data have been already
analyzed in a number of studies \cite{konstas2009social, tria2014dynamics,
  kowald2020unfairness} and they represent a sort of standard in the
music recommendation system. In particular, we focus on the Music
Listening Histories Dataset (MLHD), which contains more than 27
billion time-stamped logs extracted from Last.fm~\cite{vigliensoni17music}.
In order to build a rating matrix out of the dataset, we selected $N$ random
users (with $N=1000, 2000, 5000$)
and the top $M$ most popular artists (with $M=500, 1000$). We then
defined the entry $r_{ui}$ as the number of times the $u$-th user
listened to the $i$-th artist.\par

The empirical distributions of the ratings for different combinations of $M$ and $N$ are
displayed in~\ref{sfig:rating}, while the distributions of the similarities among
users are reported in~\ref{sfig:sim}. Both distributions turn out to be very broad.
Also the popularity of individual artists, $w_i = \sum_u r_{ui}$,
is broadly distributed (see~\ref{sfig:pop}).
This is a clear indication that users tend to give high ratings to the same set of artists,
i.e., they do not behave independently.
If users were completely independent, each one would prefer a different artist and
thus the popularity distribution would tend to be peaked.

The broad distribution of ratings suggests that to reproduce last.fm
empirical data we should consider our recommender model at the border
between multiple--item and single--item, i.e. for $\beta=1$.
We perform numerical simulations of the model dynamics
for $\beta=1$ and determine the value of the similarity bias
$\alpha$ that best fits the empirical distributions.
In this way we quantify the level of interaction among users.
Since we know that the popularity is broadly distributed,
we set $r_0=1/[N(M-1)]$, a value that for $\alpha=0$ gives a broad
popularity distribution.
We then perform numerical simulations for various
values of $\alpha$, determining the one best reproducing the
data.
The model dynamics is run for a time equal to the average
number of plays per user in the last.fm dataset.
It is worth pointing out that even if we set the average values
equal, in the real system the number of plays
largely fluctuates from user to user, while in our model each user
clicks more or less the same number of times. 

Results of numerical simulations are compared with the empirical
evidence in~\ref{fig:fit_lastfm}; a partial agreement is observed; more details about the adherence of the model to real data are reported in Appendix~\ref{Rsquared}.
We selected the $\alpha$ values such that the distributions most closely
reproduce empirical data.
These values are close to $2$ in all the cases considered, suggesting that
a non negligible amount of interaction underlies users behavior in last.fm.
These results indicate that our model is capable of reproducing the main
features of users behavior in the online platform last.fm and
allow to gauge the strength of the collaborative--filtering they are
exposed to, even if also other mechanisms we are neglecting may
  play a role. Despite our model makes significant
  improvement in modeling recommendation algorithms, we observe
  some discrepancies between numerical simulations and empirical
  data. These may arise from several features not considered in our
  schematic representation. For instance, last.fm, like many other
  online platforms, features a social network structure that
  influences the content to which users are exposed. Additionally,
  real users demonstrate temporal correlations in their behavior,
  fluctuations in music consumption, and personal preferences.
    \section{Conclusions}
Understanding the effects of recommendation algorithms in social
phenomena and their role in the polarization of opinions is a central
problem to be tackled in order to prevent the rise of
radicalization. The feedback loop these algorithms tend to establish
represents a serious threat to our society, as users are trapped in
filter bubbles where they are exposed only to content and news
confirming their past beliefs. In this paper we addressed this issue
by considering the effects of user--user collaborative--filtering, a
paradigmatic approach to algorithmic recommendations, on a group of
$N$ users allowed to repeatedly choose among $M$ different items or
opinions. Depending on the strength of the similarity bias $\alpha$,
which sets the importance the algorithm gives to choices made by
similar users, and on the magnitude of the popularity bias $\beta$,
that gauges the weight given to items with high ratings, the
phase--diagram of the system is characterized by three phases:
\begin{itemize}
    \item a \textbf{Disordered} phase for $\beta<1$ and any $\alpha$,
      where all users rate items in a completely random manner;
    \item a \textbf{Consensus} phase for $\beta>1$ and $\alpha<1$,
      where all users share the same opinion;
    \item a \textbf{Polarized} phase for $\beta>1$ and $\alpha>1$,
      where each user sticks to a given item, the system being split
      into various communities corresponding to different selected
      items.
\end{itemize}
None of these phases corresponds to viable recommendations. Indeed, in
the disordered phase the algorithm recommends just random items; in
the consensus phase it treats all users in the same way; in the
polarized phase the filter bubble problem emerges, since each user is
exposed to a single opinion. However, by looking at the transition
line between disorder and polarization we showed that it is possible
to operate the recommendation algorithm in an ``ideal'' regime,
featuring both personalized and diverse recommendations, without the
onset of filter bubbles.  On this critical line users explore more
than just a single opinion, as confirmed by a broad distribution of
the normalized ratings. We believe the approach we introduced
represents a first step toward the development of a general theory of
collaborative--filtering based recommendation algorithms, allowing to
understand their optimal regimes and their potential drawbacks.\par

We then compared our model with empirical data coming from the last.fm
music platform.  We observe broad distributions of the user ratings,
of the similarity among users and of the artists popularity. In the framework of our model, these
results can be interpreted as the outcome of a recommendation system
operating a collaborative--filtering algorithm on the critical line
$\beta=1$.  By fitting our model to the data, we were able to
qualitatively recover a broad distribution of the ratings, of
the popularity and of the similarity.
In particular, we find values of the
similarity bias $\alpha\approx 2$, which show the presence of a non
negligible effective interaction among users.\par

Clearly many realistic ingredients have not been taken into account in
the present work and deserve to be investigated. In particular, we
assumed that users have an infinite memory, so that the initial
condition is never completely forgotten and influences the properties
of the steady state. The consideration of a system where only
ratings expressed in a finite time window in the past determine the
algorithmic recommendation is an extremely interesting avenue for
future research. Additional important ingredients we neglected are the
heterogeneities in the rate of clicking, as observed in last.fm,
or possible different initial conditions for users. Understanding
whether these modifications change the overall phenomenology is
an interesting further development. Also, for a closer comparison
with empirical data from real recommender systems it would be
interesting to analyize not only the properties of the stationary state
of the system, but also the timescales needed to reach it. Finally, while this study intentionally excluded network structure to maintain model simplicity, many platforms, including last.fm \cite{asikainen2020cumulative}, are characterized by the presence of a social network, which is an important factor in the enhancement of polarization \cite{Peralta2021, doi:10.1126/sciadv.abq2044}. Future research will delve into the social interaction aspect that has been set aside in this initial model, adding depth to the understanding of the influence of these algorithms.\par

Despite the heterogeneity of users and artists
and the limitations we discussed, our model with small similarity bias
reproduces the phenomenology of last.fm users.
These results point to
the presence of a collaborative--filtering based recommendation
algorithm in the online music platform we considered and show that a
relatively simple model can capture its main
features and allow to assess the relevance of its rating and similarity biases.
%
%
%
%
\appendix
    \section{Scaling of initial conditions}
        \label{app:scaling_initial_conditions}
Let us consider a specific item that has been clicked on $k$ times at
time $t$, i.e. $r_{ui}(t)=r_0+k$. The variation of the
normalized rating $\delta \hat{r}=\hat{r}_{ui}(t+1)-\hat{r}_{ui}(t)$,
assuming it is clicked once in a time unit is,
\[
\delta \hat{r}= \dfrac{(M-1)r_0+t-k}{(Mr_0+t+1)(Mr_0+t)} \sim \dfrac{1}{M r_0}
\]
where the limit of $t \ll M r_0$ has been taken, implying that we are
dealing with the very first moments of the dynamics. Depending on the
scaling of $r_0$ with $M$ we can distinguish three possibilities
\begin{itemize}
\item if $r_0 \sim M^{-a}$ with $a>1$, $\delta \hat{r}$
  vanishes when $M \to \infty$;
  this means that in the large $M$ limit the system is not able to
  move in the phase space;
\item if $r_0 \sim M^{-a}$ with $a<1$, $\delta \hat{r}$ increases
  with $M$, being upper bounded by $1$; this means that in the
  large $M$ limit the system is able to arrive with the first click on
  the border of the phase space not being able to explore
  it all;
\item if instead $r_0 \sim M^{-1}$, then $\delta \hat{r}$ remains constant
  to a value smaller than $1$ in the large $M$ limit; this is the
  right scaling we are looking for to preserve the phenomenology of
  the system regardless of the value of $M$.
\end{itemize}
The conclusion in that in order to take the large $M$ limit the
initial condition must scale as $r_0\sim 1/M$.

\section{Master equation for the ratings}
\label{Master}

Denoting by $Q(r_{ui},t)$ the probability distribution of $r_{ui}$
at time $t$, its temporal evolution is
\[
\begin{split}
  Q(r_{ui},t+\delta t)&=Q(r_{ui}-1,t) R_{ui}(r_{ui}-1)+
  \\
  &+[1-R_{ui}(r_{ui})]Q(r_{ui},t),
\end{split}
\]
reflecting the nondecreasing evolution of the $r_{ui}$.

Expanding the l.h.s. in the continuous time limit $\delta t =1/N \to 0$ one obtains \eqref{eq:master_equation}
\[
    \dfrac{d}{dt}Q(r_{ui},t)= \frac{1}{\delta t}\left[
    \begin{multlined}[t]
        Q(r_{ui}-1,t)R_{ui}(r_{ui}-1)+\\
        -R_{ui}(r_{ui})Q(r_{ui},t),
    \end{multlined} \right]
\]

The drift coefficient, i.e., the time derivative of
the average value $\langle r_{ui}(t) \rangle$, can
be derived by
writing
\[
\dfrac{d \langle r_{ui}(t) \rangle }{dt}=\sum_{r_{ui}=r_0}^{\infty}
r_{ui} \dfrac{d}{dt}Q(r_{ui},t)
\] 
and inserting \eqref{eq:master_equation} into it
\begin{align*}
  \dfrac{d \langle r_{ui}(t) \rangle }{dt}&=\\
  &=\begin{aligned}[t]
        &\frac{1}{\delta t}\sum_{r_{ui}=r_0}^{\infty} r_{ui} [Q(r_{ui}-1,t) R_{ui}(r_{ui}-1)+\\
        &-Q(r_{ui},t) R_{ui}(r_{ui}),t)]=
  \end{aligned}\\
  &=\frac{1}{\delta t}\sum_{r_{ui}=r_0}^{\infty} [(r_{ui}+1)-r_{ui}] Q(r_{ui},t) R_{ui}(r_{ui})
  =\\
  &=\frac{1}{\delta t} \sum_{r_{ui}=r_0}^{\infty} r_{ui} Q(r_{ui},t) R_{ui}(r_{ui})=
  \frac{1}{\delta t} \langle R_{ui}\rangle =\\
  &=N \langle R_{ui}\rangle,
\end{align*}
where we have set $Q(r_0-1,t) R(r_0-1)=0$.

So far we have considered the values of the ratings. If we want to
consider the normalized ratings
$\hat{r}_{ui}(t) \approx r_{ui}(t)/(t+Mr_0)$ the drift is given by
{\small
  \begin{gather*}
    \hat{\nu}_{ui}=\dfrac{d \langle \hat{r}_{ui} \rangle }{dt} \approx
    \dfrac{d}{dt} \left(\dfrac{\langle r_{ui}
      \rangle}{t+Mr_0}\right)=\dfrac{1}{t+Mr_0}\dfrac{d \langle r_{ui}
      \rangle }{dt}-\dfrac{ \langle r_{ui}
      \rangle}{(t+Mr_0)^2}.
  \end{gather*}
}	
\section{Stationary solutions}
\label{Stationary}

The system of differential equations~\eqref{eq:drift},
at stationarity reads
\begin{equation}
  \hat{r}_{ui}=N R_{ui}=\sum_{v=1}^N \dfrac{s_{uv}^{\alpha}(t)}{\sum_w
    s_{uw}^{\alpha}(t)} \dfrac{\hat{r}_{vi}^{\beta}(t)}{\sum_j
    \hat{r}_{vj}^{\beta}(t)}.
  \label{eq:eqsolutions}
\end{equation}
The solution of these $N \times M$ equations gives the stationary
states of the model.
	
The configurations fulfilling this requirement are:\\
	
\textbf{Disorder solution}: under this condition all normalized
ratings are equal to $\hat{r}_{ui}=1/M$ for large times,
since all items are rated on average the same number of
times. Similarities are then all equal to $1$, thus
\eqref{eq:eqsolutions} is fulfilled for any value of $u$ and $i$
\[
\dfrac{1}{M}=\sum_{v=1}^N \dfrac{1}{N} \dfrac{M^{-\beta}}{M \cdot M^{-\beta}}=\dfrac{1}{M}.
\]\\
In this phase the distribution of normalized ratings is a delta function centered
around $\hat{r}_{ui}=1/M$, while the distribution of similarities is a delta function
centered around $s_{uv}=1/M$.
	
\textbf{Consensus solution}: only one item is clicked on in the limit
$t\gg 1$, then $\hat{r}_{ui}=1$ for some $i$, and
$\hat{r}_{uj}=0$, $\forall j \ne i$, and this holds for any user.
Since also in this case all similarities are equal to $1$,
\eqref{eq:eqsolutions} reads, for item $i$,
\[
1=\sum_{v=1}^N \dfrac{1}{N} \dfrac{1^{\beta}}{1^{\beta}+(M-1) \cdot 0^{\beta}}=1,
\]
while for the items $j \neq i$
\[
0=\sum_{v=1}^N \dfrac{1}{N} \dfrac{0^{\beta}}{1^{\beta}+(M-1) \cdot 0^{\beta}}=0.
\]
Hence the stationarity condition is satisfied for any $u$ and $i$.
In this phase, the distribution of the normalized ratings is
the superposition of two delta functions, one centered in 0 (weight $1-1/M$)
and one centered in 1 (weight $1/M$).
The distribution of similarities is a delta function centered in 1.

\textbf{Polarization solution}: only one item is clicked on by each user
in the limit $t \gg 1$,
but it is not the same for all users. Let us assume
that for $K$ users $\hat{r}_{u i_1}=1$ and
$\hat{r}_{u j}=0$ for $j \neq i_1$,
while for the other $N-K$ users
$\hat{r}_{u i_2}=1$ and $\hat{r}_{uj}=0$ for $j \neq i_2$.
In this case the similarity is $1$ for users rating the same item,
while it is $0$ otherwise.
It is simple to check that \eqref{eq:eqsolutions} is satisfied.
It is also easy to check that the stationarity condition
is satisfied by any possible polarized configuration
(more than two groups, of any size).
In this phase, the distribution of normalized ratings is again
the superposition of two delta functions, centered in 0 and 1,
as for the consensus solution.
Also the distribution of similarities is given by the superposition
of a delta centered in 0 and one centered in 1, with weights depending
on the distribution of group size.\\

\textbf{Other solutions}: In principle, 
there exist also other stationary solutions.
These are all configurations
where multiple--items are rated by a single user exactly in the same
proportion.
For example, $m<M$ items are equally rated by user $u$, i.e.,
$\hat{r}_{u i_1}=\dots=\hat{r}_{u  i_m}=1/m$ for $m$ items,
while $r_{uj}=0$ for $j \neq i_1, \dots, i_m$.
These solutions are stationary both if the $m$ items are the same for all
users and if they differ for different users.
This holds for any value of $m$ (for $m=1$ we have single-item solutions,
while for $m=M$ we have disorder).
In the next Appendix it is shown that these solutions are unstable
for any values of the parameters $\beta$ and $\alpha$.

\section{The transition from multiple--item to single--item}
\label{DisorderSingle}

In this Appendix we present explicit calculations about the
\emph{multiple--item} to \emph{single--item} transition occurring as a
function of $\beta$, for any value of $\alpha$.
	 
\subsection{The multiple--item solution}

Let us start from the case $\alpha=\infty$. Plugging the disorder
solution \eqref{eq:disorder_sol} into the expression~\eqref{eq:drift}
we obtain, apart from the factor $\frac{1}{t+Mr_0}$
\begin{eqnarray}
\frac{d\hat{r}_{ui}}{dt}=\frac{d \epsilon_{ui}}{dt} &
\propto &\frac{\ton*{ \dfrac{1}{M} +
    \epsilon_{ui}}^{\beta}}{\sum_j\ton*{ \dfrac{1}{M} + \epsilon_{uj}
  }^\beta}-\left(\dfrac{1}{M} + \epsilon_{ui}\right)\\
&\propto &\frac{\ton*{ 1 +
    M\epsilon_{ui}}^{\beta}}{\sum_j\ton*{ 1 + M\epsilon_{uj}
  }^\beta}-\left(\dfrac{1}{M} +\epsilon_{ui}\right).
\end{eqnarray}

Expanding for $M \epsilon_{ui} \ll 1$
leads to
\[
\frac{d\epsilon_{ui}}{dt}
\propto\frac{\ton*{ 1 + \beta
    M\epsilon_{ui}}}{\sum_j\ton*{ 1 + \beta M\epsilon_{uj}
}}-\left(\dfrac{1}{M} + \epsilon_{ui}\right)
\]
Since $\sum_j \epsilon_{uj}=0$ (see \eqref{sumrule})
then $\sum_j (1 + \beta M\epsilon_{uj})=M$ and hence
\[
\frac{d\epsilon_{ui}}{dt}
\propto \dfrac{1}{M}( 1 + \beta
M\epsilon_{ui})-\dfrac{1}{M}(1 +
M\epsilon_{ui})=(\beta-1)\epsilon_{ui}
\]
that is \eqref{stab}. 

If $\alpha$ is finite, the similarities do
not cancel from \eqref{eq:drift},
and we should consider all of them. We now show that similarity
terms are equal to $1$ up to corrections of the second order in
$\epsilon_{ui}$, which we can safely neglect for any $\alpha>0$.
In this case the situation is substantially equivalent to
the case $\alpha=0$, where all similarities are exactly equal to $1$.
		
Let us recall the expression of the cosine similarity
\[
s_{uv}=\dfrac{\sum_i \hat{r}_{ui} \hat{r}_{vi}}{\sqrt{\sum_i
    \hat{r}_{ui}^2}\sqrt{\sum_i \hat{r}_{vi}^2}}
\]
and plug into it the ratings corresponding to the disorder
solution~\eqref{eq:disorder_sol}
\[
s_{uv}=\dfrac{\sum_i \left( \frac{1}{M} + \epsilon_{ui} \right) \left(
  \frac{1}{M} + \epsilon_{vi} \right)}{\sqrt{\sum_i \left( \frac{1}{M}
    + \epsilon_{ui} \right)^2} \sqrt{\sum_i \left( \frac{1}{M} +
    \epsilon_{vi} \right)^2}}
\] 
Multiplying by $M^2$ both the numerator and the denominator and
expanding the terms in the square roots at denominator (which is
equivalent to neglect terms of order $O(\epsilon^2)$) we can rewrite
\begin{gather*}
  s_{uv} \approx \dfrac{\sum_i (1+M
    \epsilon_{ui})(1+M \epsilon_{vi})}{\sqrt{\sum_i (1+2M
      \epsilon_{ui})}\sqrt{\sum_i (1+2M
      \epsilon_{vi})}}=\\ =\dfrac{\sum_i (1+M \epsilon_{ui}+M
    \epsilon_{vi}+M^2 \epsilon_{ui}\epsilon_{vi})}{\sqrt{\sum_i (1+2M
      \epsilon_{ui})}\sqrt{\sum_i (1+2M \epsilon_{vi})}}.
\end{gather*}
Exploiting once again the fact that
$\sum_i \epsilon_{ui}=\sum_i \epsilon_{vi}=0$ we finally obtain that
\[
s_{uv} \approx \dfrac{M+M^2 \sum_i \epsilon_{ui}\epsilon_{vi}}{M}=1+M
\sum_i \epsilon_{ui}\epsilon_{vi}
\]
thus confirming the fact that $s_{uv}=1-O(M^2 \epsilon^2)$;
we specify the minus sign to remind that similarities
cannot be larger than $1$. Note that, for this reason,
the term $\sum_i \epsilon_{ui}\epsilon_{vi}$ is always negative.

To show what happens if similarities can be put equal to $1$, for
simplicity let us consider directly the case $\alpha=0$ where
similarities are exactly equal to $1$.
In this case the expression~\eqref{eq:drift} reduces to
\begin{align*}
\frac{d\hat{r}_{ui}}{dt} &\approx
\frac{1}{t+Mr_0}\qua*{\dfrac{1}{N}\sum_v\frac{\hat{r}_{vi}^{\beta}}{\sum_j\hat{r}_{vj}^{\beta}}-\hat{r}_{ui}}\approx\\ &\approx\frac{1}{t+Mr_0}\qua*{\dfrac{1}{N}
  \sum_v\frac{\ton*{ \dfrac{1}{M} +
      \epsilon_{ui}}^{\beta}}{\sum_j\ton*{ \dfrac{1}{M} +
      \epsilon_{uj} }^\beta}-\left(\dfrac{1}{M} +
  \epsilon_{ui}\right)}.
\end{align*}
Following the same procedure as in the case
$\alpha=\infty$, we can simplify this expression obtaining
\[
\frac{d\hat{r}_{ui}}{dt} = \frac{d\epsilon_{ui}}{dt} \propto \dfrac{1}{N} \beta \sum_v
\epsilon_{vi}-\epsilon_{ui}=\beta \langle \epsilon_{vi} \rangle
-\epsilon_{ui}
\]
where $\langle \cdot \rangle= \frac{1}{N} \sum_v$.

Reapplying the average over users we obtain
\[
\frac{d \langle \epsilon_{ui}\rangle}{dt} =\langle \beta
\langle \epsilon_{vi} \rangle -\epsilon_{ui}\rangle =(\beta-1)
\langle \epsilon_{vi} \rangle.
\]

Similarly to the case $\alpha=\infty$, in which we had only one user,
this expression means that when $\beta<1$, the items which are on
average underrated would tend to increase while items which are on
average overrated would tend to decrease, ensuring the stability of
the disorder solution. When $\beta>1$ the situation is reversed and
thus the disorder solution is unstable.

\subsection{The single--item solution}

Let us now consider a configuration close to a single-item solution,
i.e. a situation in which the normalized ratings are
as in \eqref{eq:singleitemsol}.
		

Let us start from the case $\alpha=\infty$ for simplicity.
Plugging \eqref{eq:singleitemsol} into \eqref{eq:drift} yields
\begin{align*}
    -\frac{d\epsilon}{dt} &= \frac{1}{t+Mr_0}\left[\frac{(1-\epsilon)^{\beta}}{(1-\epsilon)^{\beta}+(M-1)\frac{\epsilon^\beta}{(M-1)^{\beta}}}-(1-\epsilon)\right]= \\
    &=\dfrac{1}{t+Mr_0}\left[\dfrac{(1-\epsilon)^{\beta}-(1-\epsilon)^{\beta+1}-\dfrac{\epsilon^\beta(1-\epsilon)}{(M-1)^{\beta-1}}}{(1-\epsilon)^{\beta}+\dfrac{\epsilon^\beta}{(M-1)^{\beta-1}}}\right].
\end{align*}
Neglecting the prefactor and the denominator, which are positive,
stability requires the numerator to be positive, i.e.,
\[
(1-\epsilon)^{\beta}-(1-\epsilon)^{\beta+1}-\dfrac{\epsilon^\beta(1-\epsilon)}{(M-1)^{\beta-1}}>0.
\]
Some straightforward algebra leads to
\[
\left[(M-1)\left(\dfrac{1}{\epsilon}-1\right)\right]^{\beta-1}>1
\]
Raising to the power $1/(\beta-1)$, the sign of the inequality is conserved if $\beta>1$,
and thus
\[
(M-1)\left(\dfrac{1}{\epsilon}-1\right)>1 \Rightarrow \epsilon<\dfrac{M-1}{M}.
\]
Hence for small $\epsilon$ the single-item solution is stable if $\beta>1$.

On the other hand, if $\beta<1$, the sign of the
inequality is reversed and thus we obtain
\[
(M-1)\left(\dfrac{1}{\epsilon}-1\right)<1 \iff \epsilon>\dfrac{M-1}{M}.
\]
In such a case for small $\epsilon$ the single-item solution is unstable.

A similar computation can be carried out for the drift of the least
rated items $j \neq i$; it is easy to obtain that
$d\hat{r}_{uj}/dt<0$ for $\beta>1$ as long as $\epsilon<(M-1)/M$,
while $d\hat{r}_{uj}/dt>0$ for $\beta<1$ under the same assumption.
All these results indicate that the multiple--item solution is stable
for $\beta<1$, while the single-item solution is the stable one
for $\beta>1$.

Considering the case with finite $\alpha$ does not change much the reasoning.
The only difference would be to consider explicitly the
similarity terms in \eqref{eq:drift}; we can get rid of
them considering a consensus single-item configuration, i.e. where
all users have similarity very close to $1$; the computations then
trivially reduces to the case of considering one only user when
$\alpha=\infty$.

Finally, we look also at the stability of other possible
solutions of \eqref{eq:eqsolutions}, those in which $m<M$ items are
rated in the same proportion:
\[
\begin{cases}
  \hat{r}_{ui} \sim \dfrac{1}{m}+\epsilon_{ui} \qquad \text{for $m$ items} \\
  \hat{r}_{uj}\sim \epsilon_{uj} \qquad \text{for $M-m$ items}
\end{cases}
\]
Performing computations similar to those previously explained, it is possible to realize that these solutions are neither stable for
$\beta<1$, because the large ratings would decrease while the
small ones would increase (deviations are suppressed), nor for
$\beta>1$, because the large ones would increase while the small
ones would decrease (deviations are amplified).

\section{The transition from consensus to polarization}
\label{ConsensusPolarization}

In this Appendix we present explicit calculations about
the transition between consensus and polarization occurring,
as a function of $\alpha$, for $\beta>1$.

Let us consider two users $u$ and $v$ who are close to the single-item state.
Each of them is described by \eqref{eq:singleitemsol}, but
with different selected items: $i_u \ne i_v$.
The similarity between these two users is, in the limit $M\epsilon \ll 1$
\begin{eqnarray}
  s_{uv}&=&\dfrac{2(M-1)\epsilon-M
    \epsilon^2}{1-2\epsilon+M\epsilon^2}
    \approx  \dfrac{2(M-1)\epsilon}{1-2\epsilon}=O(M\epsilon).
\end{eqnarray}

Let us consider $\beta=\infty$.
If $K$ users are close to the single-item state with item $i_u$ selected,
while the remaining $N-K$ users have selected item $i_v$,
the expression of the drift is
\begin{equation}
  \frac{d\hat{r}_{ui}}{dt}=\dfrac{1}{t+Mr_0} \left[ \dfrac{\sum_v s_{uv}^{\alpha}
      \delta_{i,i_u} }{\sum_w s_{uw}^{\alpha}}-\hat{r}_{ui} \right].
		\label{eq:drift_betainfinite}
\end{equation}
Since similarities between users in the same group are
$s_{uv} \approx 1$,
while similarities between users of different groups are
$s_{uv} \approx M \epsilon$ we can write
\[
\frac{d\hat{r}_{ui_u}}{dt} \approx \dfrac{1}{t+Mr_0} \left[ \dfrac{K }{K+(N-K)
    (M\epsilon)^{\alpha}}-(1-\epsilon) \right].
\] 
The sign of the drift coefficient is thus related to the sign of the
following expression:
\begin{eqnarray}
  & &  K-(1-\epsilon)[K+(N-K)(M\epsilon)^{\alpha}]= \\
  &=&-(N-K)(M\epsilon)^{\alpha}+K\epsilon+(N-K)M^\alpha\epsilon^{\alpha+1}.
  \label{eq:drift_alfa}
\end{eqnarray}

If $\alpha \in [0,1)$, taking the limit
  $\epsilon \to 0$, the dominant term is the one of order $\epsilon^{\alpha}$,
and $d\hat{r}_{ui_u}/dt<0$.
So in this case the rating of the most rated item would tend to decrease and the
polarized solution is not stable.
On the contrary, if $\alpha > 1$, the dominant term is the one of order $\epsilon$,
so that $d\hat{r}_{ui_u}/dt>0$
and the polarized solution is stable.
	
At the same time, if one considers $j \neq i_u$,
an analogous computation gives,
in the limit of small $\epsilon$, $d\hat{r}_{uj}/dt>0$ for $\alpha<1$
(instability of the polarized solution) $d\hat{r}_{uj}/dt<0$ for $\alpha>1$
(stability of the polarized solution).
In particular, for $\alpha=1$
the drift is still negative since it is proportional to
\[
-(N-K)\epsilon+K\epsilon+(N-K)\epsilon^{2}=-N \epsilon+(N-K)\epsilon^{2}
\]
which is a negative quantity if $\epsilon<1$. Thus we can conclude
that at the critical point $\alpha=1$, the basin of attraction of the
polarized solutions is still null.
A very rough estimate of the
amplitude of the basin of attraction can be obtained from the expression
\eqref{eq:drift_alfa} by neglecting the higher order term
$(N-K)\epsilon^{\alpha+1}$, thus obtaining a solvable inequality which
states that the drift points toward the polarized solutions at least
until
\[
\epsilon < \left( \frac{K}{N-K} \right)^{\frac{1}{\alpha-1}}.
\] 
The same argument applies to any possible polarized
solution, i.e. every value of $K$ between $1$ and $N-1$ and an
arbitrary number of groups.

For what concerns the case of generic $\beta>1$,
the only difference with \eqref{eq:drift_betainfinite} is that
the terms ${\hat{r}_{ui}^{\beta}}/{\sum_j \hat{r}_{uj}^{\beta}}$
must be considered explicitly;
but in the limit of small $\epsilon$ we have
  \begin{align*}
    \lim_{\epsilon \to 0}\dfrac{\hat{r}_{ui_u}^{\beta}}{\sum_j
      \hat{r}_{uj}^{\beta}} &= \lim_{\epsilon \to 0}
    \frac{(1-\epsilon)^{\beta}}{(1-\epsilon)^{\beta}+(M-1)
      \frac{\epsilon^\beta}{(M-1)^{\beta}}}=\\
      &=\lim_{\epsilon \to 0}1-O(\epsilon)=1\\
    \lim_{\epsilon \to 0} \dfrac{\hat{r}_{ui \neq
        i_u}^{\beta}}{\sum_j \hat{r}_{uj}^{\beta}} &= \lim_{\epsilon \to0}
    \frac{\dfrac{\epsilon^{\beta}}{(M-1)^{\beta}}}{(1-\epsilon)^{\beta}+(M-1)
      \frac{\epsilon^\beta}{(M-1)^{\beta}}}=\\
      &=\lim_{\epsilon \to 0}O(\epsilon)=0
  \end{align*}
thus we are led back to \eqref{eq:drift_betainfinite}.
As a consequence, for every $\beta>1$, we
obtain that the critical value of the transition is $\alpha=1$.

\section{The Order Parameters}
\label{OrderParameters}

We define the item fluctuations $V_i$ as
\begin{equation}
  V_i = \frac{M^2}{M-1}\frac{1}{N}\sum_u \qua*{
    \dfrac{1}{M}\sum_i \hat{r}_{ui}^2-\ton*{\dfrac{1}{M}\sum_i
      \hat{r}_{ui}}^2}
  \label{eq:item_fluctuation}
\end{equation} 
which is the variance of users' normalized rating vectors
$\vec{\hat{r}}_u=(\hat{r}_{u1}, \hat{r}_{u2},\ldots, \hat{r}_{uM})$
averaged over all users. The factor $M^2/(M-1)$ ensures that
$V_i$ varies in the interval $[0, 1]$.
In the disordered phase it holds $\hat{r}_{ui} = 1/M$,
and so we have $V_i = 0$.
Conversely, when users stick to a single item the normalized
ratings satisfy $\hat{r}_{ui}=1$ and $\hat{r}_{uj}=0$ for $j\neq i$,
thus giving $V_i=1$.

Analogously, we introduce the user fluctuations $V_u$ as
\begin{equation}
  V_u = \frac{M^{2(N-1)}}{M^{N-1}-1}\frac{1}{M}\sum_i \qua*{
    \dfrac{1}{N}\sum_u \hat{r}_{ui}^2-\ton*{\dfrac{1}{N}\sum_u
      \hat{r}_{ui}}^2}
  \label{eq:user_fluctuation}
\end{equation} 
which is the variance of the items' normalized rating vectors
$\vec{\hat{r}}_i=(\hat{r}_{1i}, \hat{r}_{2i},\ldots, \hat{r}_{Ni})$
averaged over all users and normalized to be in $[0, 1]$.
It is easy to see that both in disorder and consensus it holds $V_u = 0$,
while in polarized states $V_u>0$, the precise value depending on the size
distribution of the groups.
In terms of these order parameters the three phases are identified by:
\begin{itemize}
\item \textbf{Disorder:} $V_i=0$ and $V_u=0$
\item \textbf{Consensus:} $V_i>0$ and $V_u=0$
\item \textbf{Polarization:} $V_i>0$ and $V_u>0$
\end{itemize}

\begin{figure*}
  \centering
  \begin{subfigure}[b]{0.475\textwidth}
    \centering
    \includegraphics[width=\textwidth]{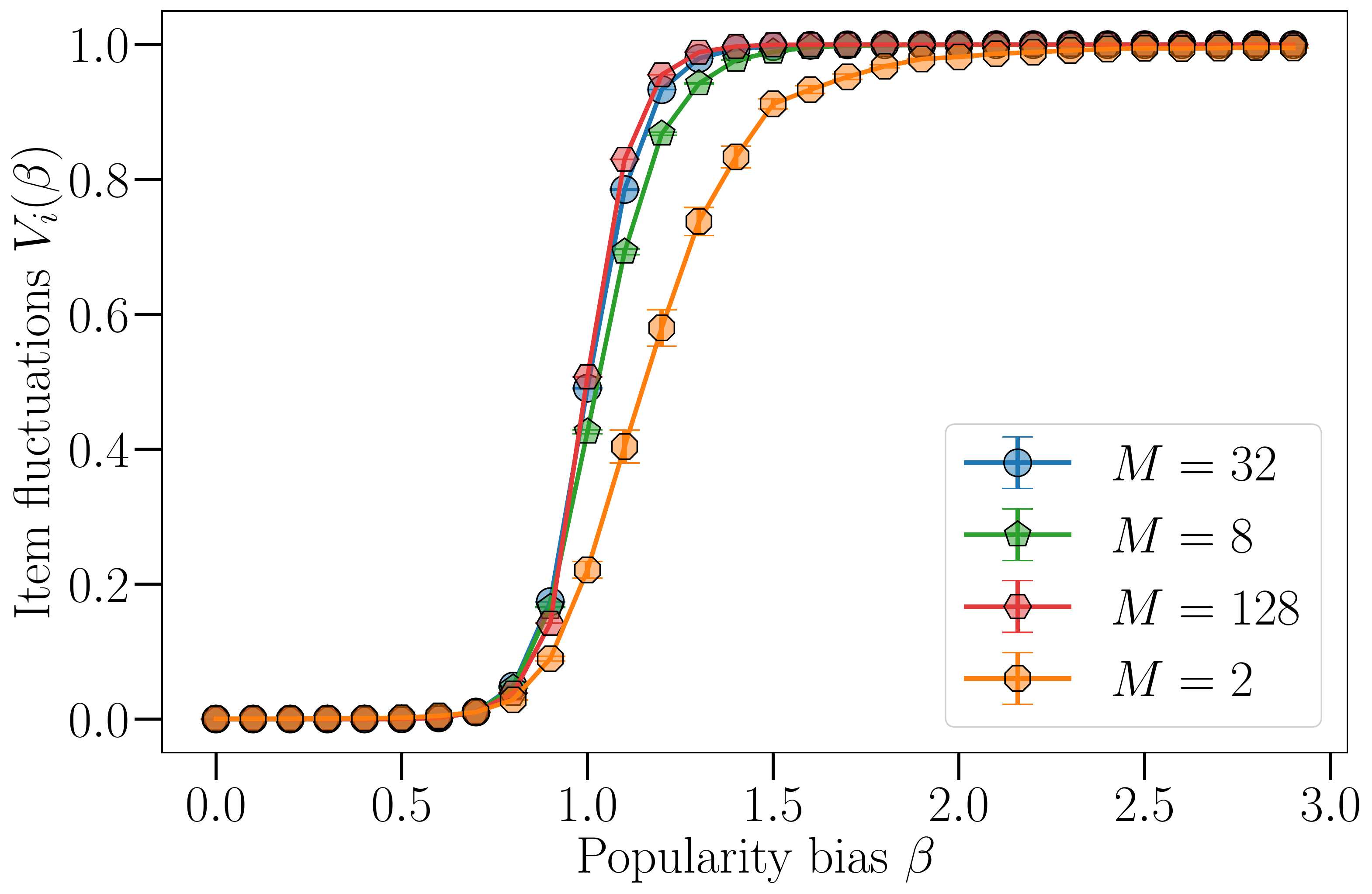}
    \caption{$N=2$}     
    \label{fig:Vi_M}
  \end{subfigure}
  \hfill
  \begin{subfigure}[b]{0.475\textwidth}  
    \centering 
    \includegraphics[width=\textwidth]{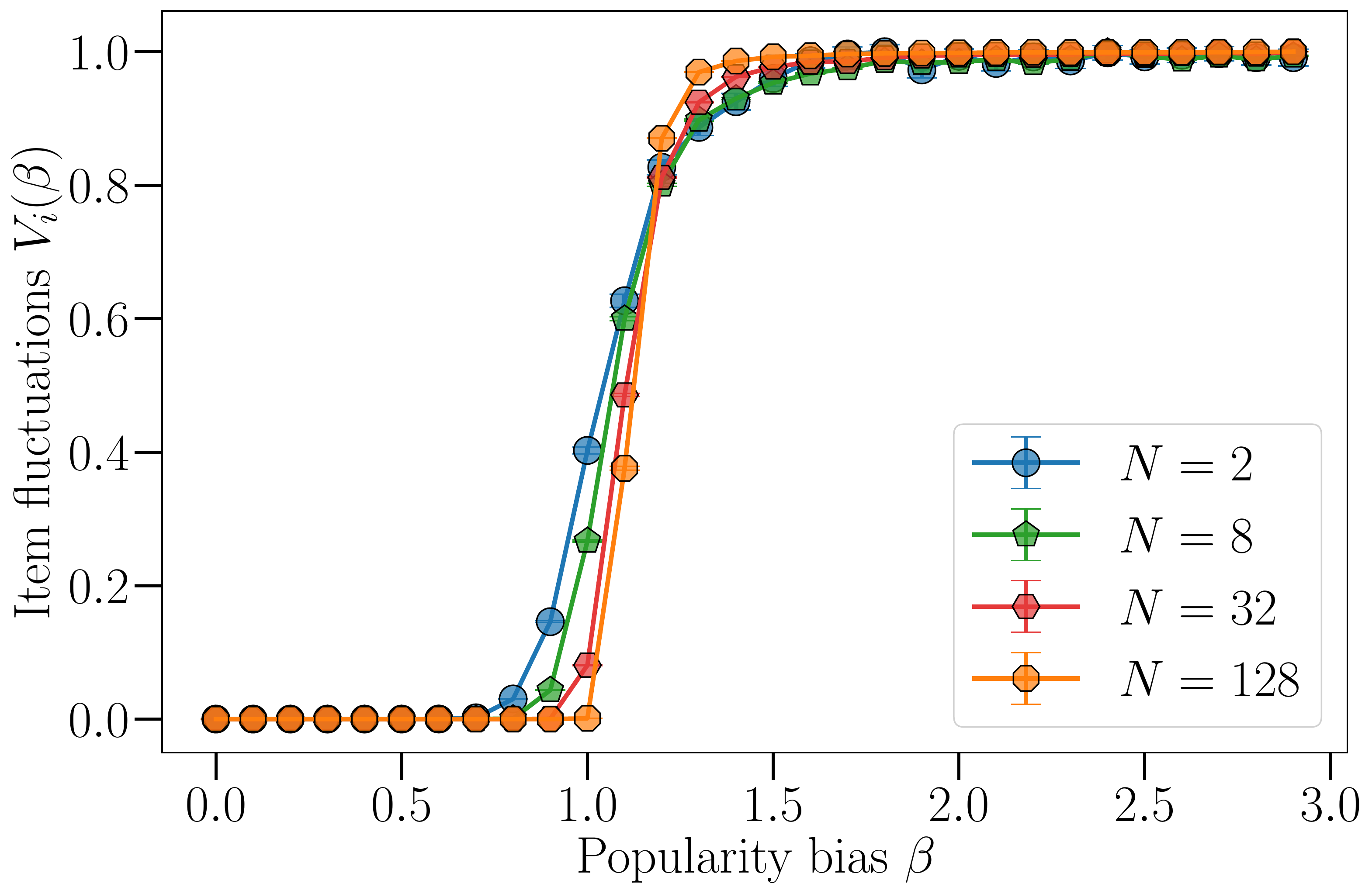}\\
    \caption{$M=8$}    
    \label{fig:Vi_N}
  \end{subfigure}
  \begin{subfigure}[b]{0.475\textwidth}   
    \centering 
    \includegraphics[width=\textwidth]{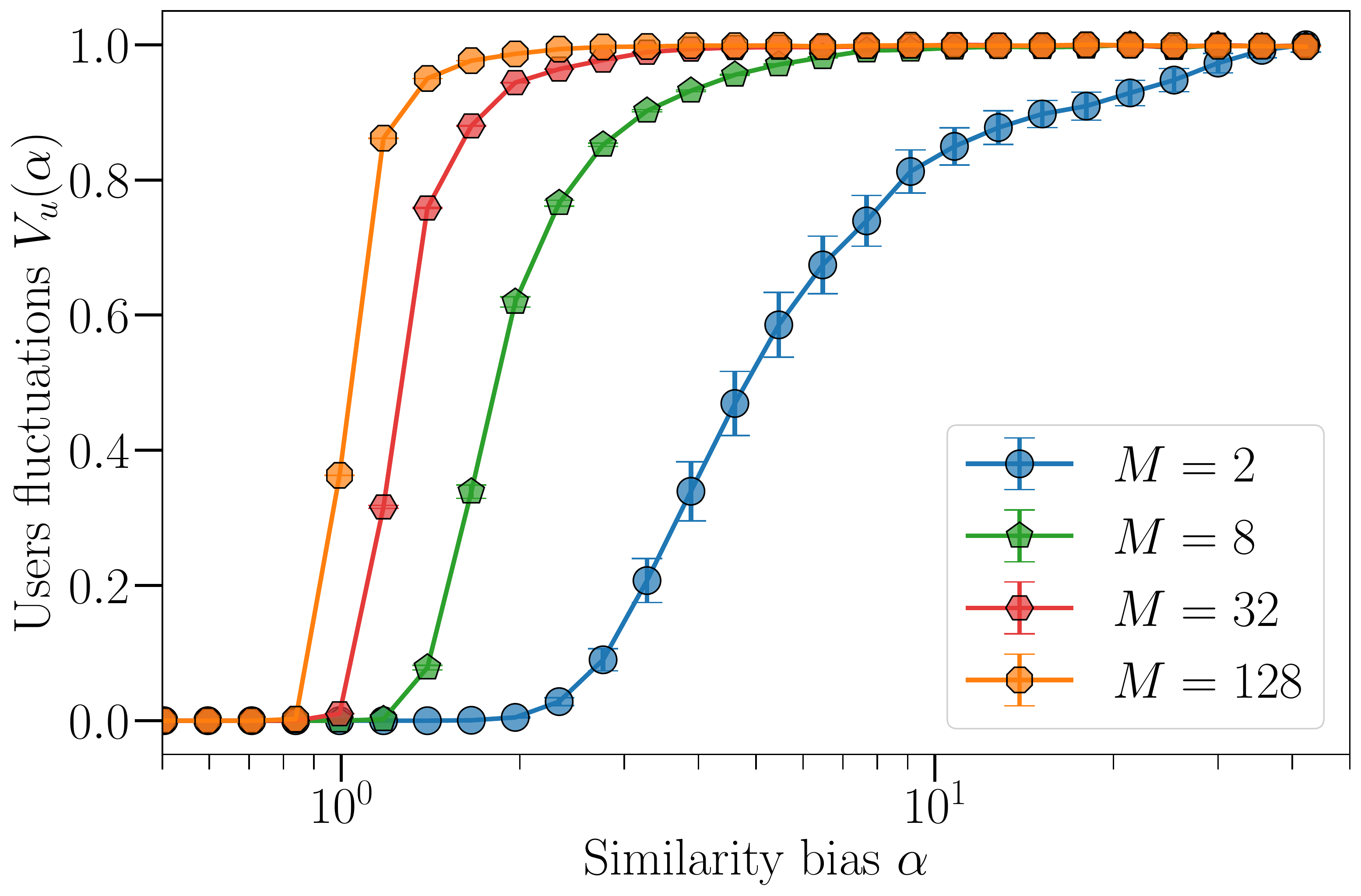}
    \caption{$N=8$}      
    \label{fig:Vu_M}
  \end{subfigure}
  \hfill
  \begin{subfigure}[b]{0.475\textwidth}   
    \centering 
    \includegraphics[width=\textwidth]{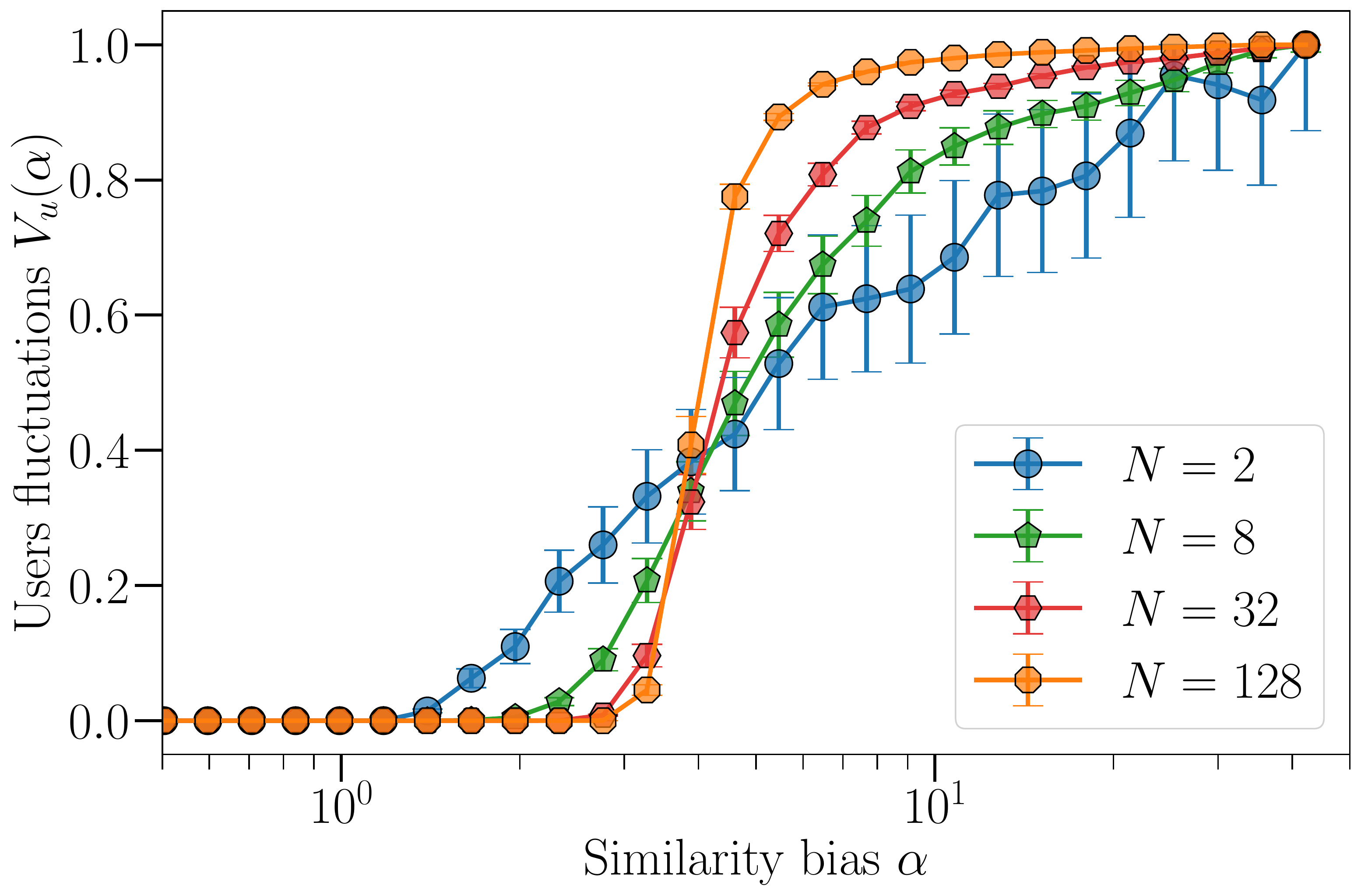}
    \caption{$M=2$}   
    \label{fig:Vu_N}
  \end{subfigure}
  \caption{\textbf{Order Parameters and Phase Transitions:} Panels
    \subref{fig:Vi_M} and \subref{fig:Vi_N} show the order parameter
    $V_i$ for the multiple--item to single--item transition as a
    function of $\beta$, for $\alpha=5$. Curves are computed averaging
    over 1000 realizations of the dynamics.  The increase of $M$ or $N$
    has no effect on the critical value $\beta_c$, but simply sharpens
    the transition.  Panels \subref{fig:Vu_M} and \subref{fig:Vu_N}
    report the behavior of the order parameter $V_u$ for the
    consensus--polarization transition. It turns out that increasing
    $M$ moves the transition towards $\alpha_c=1$.
    Increasing $N$ for fixed $M$ has little effect with regard to
    the location of the transition while it makes the transition
    sharper. This indicates that at fixed $M$, also in the limit $N \to
    \infty$ we can see the transition only after a value
    $\alpha_c^{\rm eff}>1$.}
  \label{fig:orderparameters}
\end{figure*}

In order to validate the phase--diagram discussed above, we perform
computer simulations of the model dynamics, focusing on the order
parameters just defined. We run the simulations
for a time equal to $T=1000 M$. 

First we look at the multiple--item to single--item transition, occurring as
a function of $\beta$. We show in \ref{fig:Vi_M} and~\ref{fig:Vi_N}
the order parameter $V_i$ as a function of $\beta$ for $\alpha=5$ and different
combinations of $M$ and $N$. As
expected we observe a transition in $\beta_c=1$ that becomes increasingly
sharper as the system size grows.

\ref{fig:Vu_M} shows how the order parameter $V_u$ depends on
$\alpha$ for fixed $N$. As $M$ is increased, the transition becomes
sharper and sharper and moves to the critical point $\alpha_c$. For fixed $M$ the transition is
instead observed at $\alpha>\alpha_c=1$ and becomes sharper with
growing $N$ (\ref{fig:Vu_N}).  This indicates that our
predictions are accurate when $M$ is sufficiently large, a situation
always occurring in real recommendation systems.

\section{The Polya Urn model}
\label{BetaDirichlet}

The Polya Urn model considers an urn where initially there are
$\boldsymbol{r_0}=\{r_{10}, \dots, r_{M0}\}$ balls of $M$ different
colors; at each time step a ball is randomly extracted from the urn,
then it is reinserted with others $S$ balls of the same color.
	
From the general results of the Polya Urn
model~\cite{polya1930quelques} it can be obtained that the
distribution of the normalized ratings follows a Multivariate Beta
distribution:
\[
P(\vec{\hat{r}})=\frac{\prod_i
  r_i^{\frac{r^{(0)}_i}{S}-1}}{\mathcal{D}\ton*{\frac{\vec{r_0}}{S}}}=\frac{\prod_i
  \hat{r}_i^{r_0-1}}{\mathcal{D}\ton*{\vec{r_0}}},
\]
where $\mathcal{D}$ is the multivariate Beta Function, $S=1$ since
each rating is increased by one subsequently to a click, and we focus
on the case of uniform initial conditions $r^{(0)}_i=r_0$ for any $i$.
This distribution is defined on the standard $(M-1)$ symplex by the
constraint of the normalization of the ratings.

These initial conditions also imply that each rating $\hat{r}_i$ is
statistically equivalent to the others; thus it is sufficient to
obtain the marginal distribution of a single rating to have the
complete statistics of the system.

In particular, the marginal one-dimensional distribution of a
Multivariate Beta Function
gives a Beta distribution $\beta(x_i; a_i, \sum_{j \neq i} a_j)$,
which in our case reads:
\[
P(\hat{r}_i)=\beta[\hat{r}_i; r_0, (M-1)r_0]=\dfrac{\hat{r}_i^{r_0-1}
  (1-\hat{r}_i)^{(M-1)r_0-1}}{B[r_0, (M-1)r_0]},
\]
where $B(x,y)$ is Euler Beta function. 

Depending on the value of $r_0$ we thus have different behaviors.
For $r_0=1/(M-1)$ the
distribution of the single ratings shows an exact power-law decay:
\begin{equation}
P(\hat{r}_i)=\dfrac{\Gamma \left(\dfrac{1}{M-1} \right) }{\Gamma
  \left(\dfrac{M}{M-1} \right)} \ \ \hat{r}_i^{-(M-2)/(M-1)}.
\label{Polya1}
\end{equation}
If $r_0<1/(M-1)$ the power-law decay for small $\hat{r}_i$ is followed by a divergence in $\hat{r}_i=1$
so that the distribution is bimodal.
If instead $r_0>1/(M-1)$ then $P(\hat{r}_i)$ is truncated by an exponential cutoff.

The mapping to a Polya Urn can be used also for the case $\alpha=0$, although
in this case the matching is only approximate. Indeed, for $\alpha=0$
\eqref{eq:transition_prob} becomes
\[
R_{ui}=\frac{1}{N^2}\sum_v\dfrac{r_{vi}}{\sum_j r_{vj}}
\]
and since on average each user is updated once at each time step we can write 
\[
R_{ui}\approx\frac{1}{N^2}\sum_v\dfrac{r_{vi}}{t+Mr_0}
\approx\frac{1}{N}\dfrac{r_{ui}}{t+Mr_0},
\] 
where we have assumed that, since $\alpha=0$, all users
behave the same so that $1/N \sum_v r_{vi}\approx r_{ui}$.
Defining the variable $w_{ui}=r_{ui}/N$ we get 
\[
R_{ui}\approx\dfrac{w_{ui}}{t+Mr_0}
\]
which has the same form of \eqref{eq:transition_polya}.
As a consequence also the dynamics of the variable $w_{ui}$ is
described by a Polya Urn, but now the reinforcement parameter is $S=1/N$.
Since the normalized variables $\hat{r}_{ui}$ and
$\hat{w}_{ui}$ coincide we can then write
\begin{align*}
P(\hat{r}_{ui})&= P\ton*{\hat{w}_{ui}}=\nonumber\\
& = \frac{\hat{r}_i^{r_0N-1}\ton*{1-\hat{r}_i}^{(M-1)r_0N-1}}{B(r_0N, (M-1)r_0N)}.
\end{align*}
If we set $r_0=1/[N(M-1)]$, this expression implies a power-law decay for $\hat{r}_{ui}$, with exponent $-(M-2)/(M-1)$. As in the case $\alpha=\infty$ we considered above, other values of $r_0$ give different broad distributions.

\section{Adjusted R-Squared for real data fits}
\label{Rsquared}

In this appendix, we present the outcomes of the Adjusted R Squared analysis for the distributions of ratings, similarity, and popularity. The $\overline{R}^2$ is a measure of the extent to which the variation in the dependent variable is predictable from the independent variable. This value aids in evaluating the efficacy of a model's prediction against actual observed data.

Given the real data $(x_1, \dots, x_n)$ and their relative predictions $(z_1, \dots, z_n)$, then we define the R Square $R^2$ as:
\[
	R^2 = 1 - \dfrac{\Sigma_{res}}{\Sigma_{tot}}
\]
where $\Sigma_{res}= \sum_i (x_i - z_i)^2$ is the sum of squares of the residuals, while $\Sigma_{tot} = \sum_i (x_i - \overline{x})^2$, where $\overline{x}$ is the average of the data, is the total sum of squares (proportional to the variance of the data). 

The value of $R^2$ spans from $1$ when the data are exactly explained by the model ($\Sigma_{res}=0$), to $0$ in the case of a baseline model which always predicts $\overline{x}$. Negative values of $R^{2}$ are associated to models under the baseline.

The Adjusted R Square $\overline{R}^2$ refines this metric by counteracting the tendency of $R^2$ to rise when additional independent variables are introduced to the model, even if these variables insignificantly contribute to the explanation of the variance. The adjusted measure is obtained considering a normalization based on the number of independent variables of the model $k$ and the number of observations $n$:
\[
	\overline{R^2} = 1- (1-R^2) \dfrac{n-1}{n-k-1}
\] 
This adjustment addresses overfitting, as an excessive number of extraneous variables leads to a decrease in $\overline{R}^2$. The resulting values still range from $0$ to $1$ and retain the same interpretation.

We perform this analysis in the three scenarios: N = 1000, M = 500, N = 2000, M = 1000, and N = 5000, M = 1000. Additionally, for the popularity distribution, we include a comparison with the R obtained from simulations of a standard urn model. The summarized results are presented in the tables below: \\

\begin{table}[H]
    \centering
    \captionsetup{justification=centering}
    \begin{tabular}{|c|c|c|c|c|c|}
        \hline
        N & M & $\overline{R}^2_{rat}$ & $\overline{R}^2_{sim}$ & $\overline{R}^2_{pop, CF}$ & $\overline{R}^2_{pop, urn}$ \\
        \hline
        1000 & 500 & 0.880 & 0.964 & 0.639 & 0.411\\
        2000 & 1000 & 0.909 & 0.933 & 0.726 & 0.504\\
        5000 & 1000 & 0.932 & 0.963 & 0.585 & 0.292\\
        \hline
    \end{tabular}
    \caption{Adjusted R-Squared for the distribution of the Ratings, Similarity and Popularity. In this last case, real data are also compared with the results from a standard Polya Urn model.}
\end{table}

The results reveal high values for the Adjusted R-Squared for the distributions of ratings and similarity. As for the popularity distribution, the values are slightly smaller but still within an acceptable range. Furthermore, it is noteworthy that our model outperforms the standard urn model in all cases, as evidenced by the higher $\overline{R}^2$.

\end{document}